\documentclass[aps,prb,reprint,twocolumn,superscriptaddress,showpacs]{revtex4-1}
\usepackage{amsmath}
\usepackage{amssymb}
\usepackage{braket}
\usepackage{graphicx}
\usepackage{hyperref}
\usepackage{color}

\newcommand{\abs}[1]{\left| #1 \right|}
\renewcommand{\vec}[1]{\boldsymbol{#1}}
\DeclareMathOperator{\sign}{sgn}

\begin{document}
    \title{Electrically controlled crossover between $2\pi$- and $4\pi$-Josephson
        effects through topologically-confined channels in silicene}

    \author{Daniel Frombach}
    \affiliation{Institut f\"ur Mathematische Physik, Technische Universit\"at
        Braunschweig, D-38106 Braunschweig, Germany}
    \author{Sunghun Park}
    \affiliation{Departamento de F\'{\i}sica Te\'orica de la Materia Condensada, Condensed Matter Physics Center (IFIMAC) and
Instituto Nicol\'as Cabrera, Universidad Aut\'onoma de Madrid, Spain}
         \author{Alexander Schroer}
    \affiliation{Institut f\"ur Mathematische Physik, Technische Universit\"at
        Braunschweig, D-38106 Braunschweig, Germany}
 \author{Patrik Recher}
    \affiliation{Institut f\"ur Mathematische Physik, Technische Universit\"at
        Braunschweig, D-38106 Braunschweig, Germany}
    \affiliation{Laboratory for Emerging Nanometrology Braunschweig, D-38106 
        Braunschweig, Germany}

    \begin{abstract}
   We propose a tunable topological Josephson junction in silicene where electrostatic gates could switch between a trivial and a topological junction. These aspects are a consequence of a tunable phase transition of the topologically confined valley-chiral states from a spin-degenerate to a spin-helical regime. We calculate the Andreev bound states in such a junction analytically using a low-energy approximation to the tight-binding model of silicene in proximity to s-wave superconductors as well as numerically in the short- and long-junction regime and in the presence of intervalley scattering. Combining topologically trivial and non-trivial regions, we show how intervalley scattering can be effectively switched on and off within the Josephson junction. This constitutes a topological Josephson junction with an electrically tunable quasiparticle poisoning source.
    \end{abstract}

\date{\today}
    \maketitle

\section{Introduction}
The fractional ($4\pi$) Josephson effect is one of the key signatures in junctions between topological superconductors hosting Majorana bound states (MBS) \cite{Kitaev2001,FuKane2009JosephsonJunction, CriticalSupercurrentBeenakker, Crepin2014}. MBS have been proposed as the fundamental building blocks for topological quantum computation \cite{Nayak2008, Pachos2012}.  The 4$\pi$-Josephson effect is inherently a non-equilibrium effect which sensitively depends on a protected crossing of many-particle states of opposite fermion parity. 
In the absence of quasiparticle poisoning \cite{FuKane2009JosephsonJunction, Rainis2012}, the parity is conserved and the current exhibits a 4$\pi$ periodicity which is known as a fermion parity anomaly (for a review see, e.g. Ref.~\onlinecite{Kane2015}). 
However, the $4\pi$-Josephson effect could in principle also originate from a ballistic spin degenerate channel where the appearance of a crossing of Andreev bound states (ABS) at a phase difference of $\phi=\pi$ is universal \cite{Beenakker91}. In the absence of scattering and equilibration effects, the current should also exhibit a $4\pi$-Josephson effect carried by each Kramer's pair. Depending on the global fermion parity of both Kramers' pairs, the current is either $4\pi$ periodic (even global fermion parity) or $2\pi$ periodic (odd global fermion parity). This result was actually derived for the quantum spin Hall (QSH) effect samples where the superconductors cover both helical edge states, and the scattering between different Kramers' pairs is prohibited by the insulating bulk, separating the two edges of the sample \cite{Crepin2014}. It means that a ballistic Josephson junction could exhibit both, a $2\pi$- or $4\pi$-Josephson effect depending on the global fermion parity.

    \begin{figure}
        \includegraphics[width = \columnwidth]{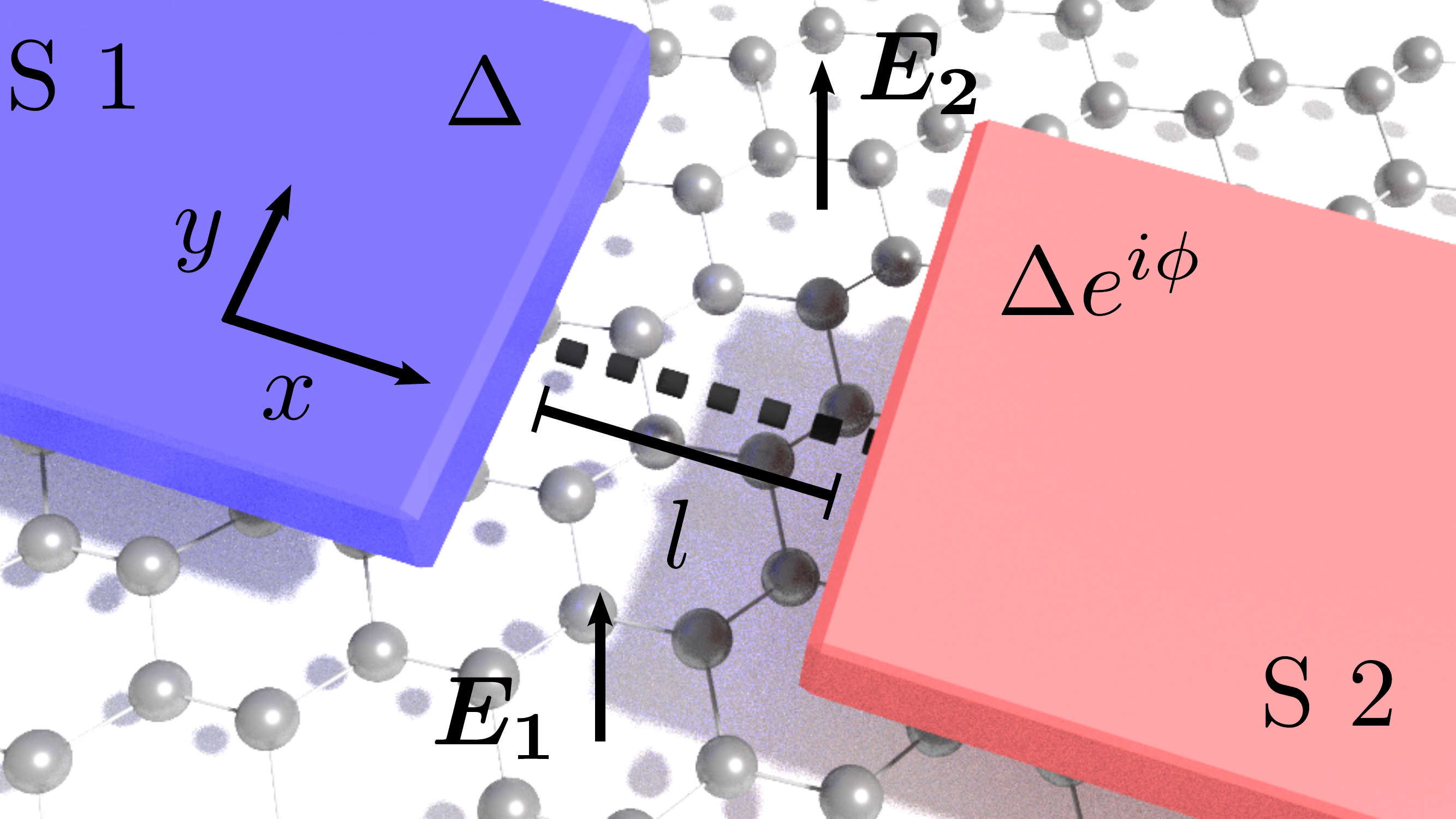}
        \caption{Topological edge states can form at the interface (gray dashed line) between two gapped regions of silicene with different applied perpendicular electric fields $ \vec{E}_1 $ and $ \vec{E}_2 $. Two superconductors (S~1 and S~2) are placed on the edge states and induce superconductivity in them via the proximity effect. The superconductors exhibit a phase difference $ \phi $ and form a Josephson junction of length $ l $ with Andreev bound states.}
        \label{fig-setup}
    \end{figure}
    
Experiments to detect the fractional Josephson effect so far have concentrated on ac-properties \cite{Rokhinson2012, Wiedenmann2016, Deacon2017} and it could be interesting to think about probably simpler dc-measurements.  It was theoretically shown \cite{CriticalSupercurrentBeenakker} that the critical current of a Josephson junction should behave differently in the short and long junction limit regarding the difference between a $4\pi$- and $2\pi$-junction. This probe only needs a static dc measurement which is not sensitive to the dynamics of $\phi$. Therefore, to pinpoint the existence of a $4\pi$-Josephson junction, a spectacular test would be the creation of a Josephson junction that could be switched (with a knob) between a $4\pi$- and $2\pi$-situation. This is what we would like to propose and investigate in this work using the peculiarities of silicene.

We investigate a novel Josephson junction that can be tuned between a $2\pi$- and a $4\pi$-Josephson junction using the valley chiral states at a mass domain in silicene, where the sign of the mass can be tuned electrically \cite{LowEnergyModelOfSilicene, SpinHelicalEdgeStatesInSilicene, Chan14, SiliceneEdgeStates, Ezawa15, Geissler2013} due to the buckled structure of its honeycomb lattice. In addition, the presence of a sizeable spin-orbit coupling allows one to switch between valley-chiral but spin-degenerate states and a valley-chiral and spin helical regime. We analytically calculate the ABS in the short junction limit using the Bogoliubov de Gennes (BdG) equation of silicene \cite{Linder14} including the generic case of intervalley scattering. We supplement the silicene Josephson junction with numerical simulations on a square lattice by discretizing the low energy model as well as by using the full tight-binding model. The intervalley scattering affects the topologically trivial regime and the ABS develop an anticrossing around $\phi=\pi$, whereas the crossing is protected in the valley-chiral and spin-helical regime by the same symmetries as for the single edge of a QSH system. This insight can be exploited to create a controllable poisoning source by constructing regions of different mass terms such that a dissipative helical channel---attached to an additional normal reservoir---is coupled to another helical channel along a small part of the Josephson junction by intervalley scattering (see Fig.~\ref{poisoning}). The presence or absence of the dissipative channel can be tuned by electric gates.

In the absence of intervalley scattering the transition between the topologically different junctions is reminiscent of the transition between a single QSH edge and a pair of QSH edges between two superconducting contacts.

\section{Model}
        We consider a sheet of silicene with two regions separated by the $ x $-axis that can be distinguished by the applied perpendicular electric field $ \vec{E}(y) $ (Fig.~\ref{fig-setup}). Two s-wave superconductors on top of the sheet induce a superconducting pairing potential in silicene via the proximity effect \cite{Linder14}. The superconductors exhibit a phase difference $ \phi $ and form a Josephson junction. At low energies, the system can be described by the Hamiltonian (setting the chemical potential to zero)
\begin{equation}
\label{eq:Hamiltonian}
H = \frac{1}{2} \int d^2 x\, \Psi^\dagger(\vec{x}) \mathcal{H}(\vec{x}) \Psi(\vec{x}),
\end{equation}       
        with   $\mathcal{H}(\vec{x}) = \mathcal{H}_0 + \mathcal{H}_{S} + \mathcal{H}_{I}$ to be described below. We use the Nambu basis
        \begin{equation}
        \label{eq:Basis}
        \Psi(\vec{x}) = 
        (\psi_{\uparrow}(\vec{x}), 
        \psi_{\downarrow}(\vec{x}),
        {\bar \psi_{\downarrow}}^\dagger(\vec{x}),
        -{\bar \psi_{\uparrow}}^\dagger(\vec{x}))^T, 
        \end{equation}
      where $
         \psi_{s}(\vec{x}) = 
         (c_{A K s}(\vec{x}),
         c_{B K s}(\vec{x}),
         c_{A K' s}(\vec{x}),
         c_{B K' s}(\vec{x}))^T$ and $ {\bar \psi}_{s}(\vec{x})$ is obtained from $\psi_{s}(\vec{x})$ by the substitution $K\leftrightarrow K'$. The operator $c_{\sigma \tau s} (\vec{x})$ annihilates an electron at position $\vec{x}$ on sublattice $ \sigma $, in valley $ \tau $ and with spin-polarization $ s $. The Hamiltonian describing the system without superconductivity and intervalley scattering is
        \begin{equation}
        \label{eq:Hamiltonian:Parts}
         \mathcal{H}_0 = -i\hbar v_F (\partial_x \rho_z \tau_z \sigma_x + \partial_y \rho_z \sigma_y) 
         + \Delta_{\textrm{SO}} \rho_z s_z \tau_z \sigma_z
         + m \rho_z \sigma_z.
        \end{equation} 
        It consists of the kinetic part (first term) with the Fermi velocity $ v_F = 5.42 \times 10^5 \, m/s $ \cite{LowEnergyModelOfSilicene}, intrinsic spin orbit interaction with $ \Delta_{\textrm{SO}} = 3.9 \textrm{ meV} $ \cite{LowEnergyModelOfSilicene} and a staggered potential  with mass $ m = 0.23 \textrm{ \AA} \times E_z $ resulting from the perpendicular electric field due to the buckled lattice structure of silicene \cite{Ezawa15, Drummond2012}. The matrices $ \sigma_i $, $ \tau_i $, $s_{i}$ and $ \rho_i $ are four sets of Pauli matrices corresponding to the sublattice subspace, the valley subspace, the electron spin and the particle-hole subspace, respectively \footnote{For definiteness, we assume $\Delta_{\textrm{SO}}>0$ in the following. If  $\Delta_{\textrm{SO}}<0$ all results remain the same if we change the basis in Eq.~(\ref{eq:Basis}): $\Psi(\vec{x}) \rightarrow U\Psi(\vec{x})$ with $U=s_x$.}. The proximity-induced s-wave superconductivity takes the form 
        \begin{equation}
         \mathcal{H}_{S} = \Delta (\cos(\phi) \rho_x + \sin(\phi) \rho_y),
        \end{equation} 
        with the superconducting pairing potential $ \Delta e^{i \phi} $. In clean silicene samples, the two valleys are independent of each other, however, in disordered samples intervalley scattering
        \begin{equation}
        \label{eq:intervalleyScattering}
         \mathcal{H}_I = \delta \rho_z \tau_x
        \end{equation} 
        of strength $ \delta $ will be present due to atomic scale impurities
        \footnote{Eq.~\eqref{eq:intervalleyScattering} is not the only possible form of intervalley scattering, but one of six time reversal symmetric and spin independent possibilities \cite{McCann_2006}. The term chosen here opens a gap in the spectrum of the spin degenerate valley chiral edge states. Important for our purposes, however, is only some kind of backscattering in the channel that would be present for any type of intervalley scattering. This would open a gap in the ABS spectrum \cite{Beenakker91}}.
        The Hamiltonian $H$ can be diagonalized $H= \frac{1}{2} \sum_n \varepsilon_n \gamma_n^{\dagger}\gamma_n$ with the creation operator
         $\gamma^\dagger_n = \int d^{2}x \,\Psi^\dagger(\vec{x}) \Lambda_{n}(\vec{x})$, where $ \Lambda_n(\vec{x})$ is
       a $16 $-component eigenspinor of the matrix $ \mathcal{H}(\vec{x})$ with eigenvalue  $\varepsilon_n$. Due to the electron-hole symmetry of the BdG-equation, $\{ \Xi,\mathcal{H}(\vec{x}) \} =0$ where $\Xi=\rho_y s_y\tau_x{\cal C}$ with ${\cal C}$ denoting the operator of complex conjugation. The solutions come in pairs $(\Lambda_n(\vec{x}),\varepsilon_n)$ and $(\Xi\Lambda_n(\vec{x}),-\varepsilon_n)$ with the corresponding operators $\gamma_{\varepsilon_n}=\gamma_{-\varepsilon_n}^{\dagger}$.
        
	\begin{figure}
	    \includegraphics[width = \columnwidth]{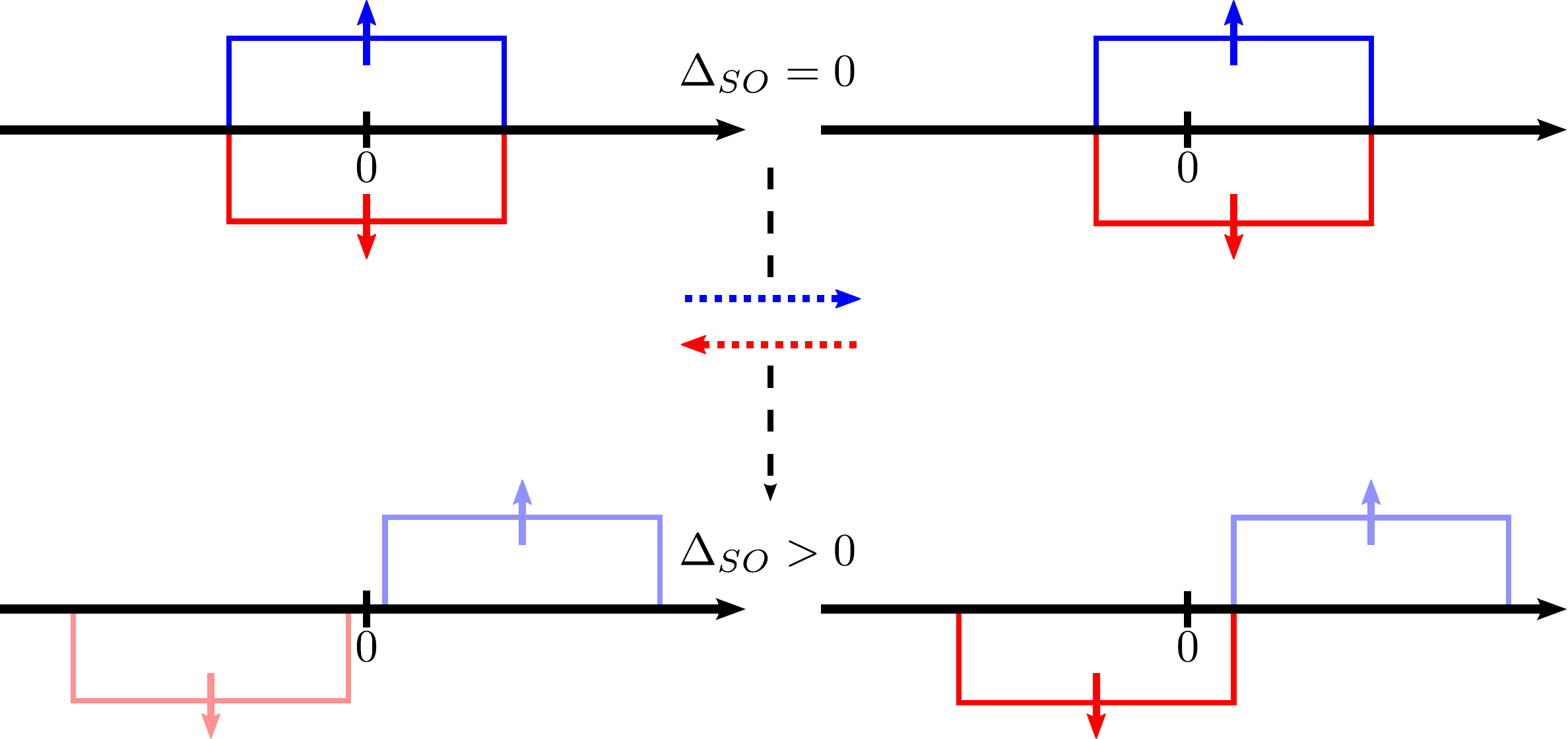}
	    \caption{Schematics of the energy gaps confining the topological channel: the energy gap sizes in both half spaces of the silicene sheet are marked on an arbitrary energy scale (horizontal arrow), forming a box. The left and right sides correspond to two different possible configurations of the mass parameters while the top and bottom halves correspond to vanishing and finite spin orbit interaction, respectively. If zero is contained in a box, the sign of the energy gap which is proportional to the Chern number changes across the interface of the two half spaces and a topological edge state exists at the interface. Since the spin in the $ z $-direction is conserved, the Chern numbers can be considered independently for each spin (blue, red). The intrinsic spin-orbit interaction strength is equal in both half spaces, so it only results in a shift of the boxes. Since the boxes for both spins get shifted in opposite directions (depicted by dashed arrows), a spin helical regime (bottom right) results in the regime $ \abs{m_i} < \Delta_{\textrm{SO}} < \abs{m_j} $, $i,j=1,2$ ($i\neq j$).}
	    \label{fig-edgeStates}
	\end{figure}

\section{Spin-helicity of valley-chiral boundary states} 	
We start by investigating $ \mathcal{H}_0 $ in the absence of superconductivity and intervalley scattering. Topologically confined edge states can be found at the interface between the two half spaces (1 with $y<0$; 2 with $y>0$) that can be distinguished by the two different perpendicular electric fields \cite{SpinHelicalEdgeStatesInSilicene, SiliceneEdgeStates} (assumed to be homogenous within each half space). The bulk dispersion of silicene in the presence of an electric field
\begin{equation}
 \varepsilon_{\vec{k}} = \pm\sqrt{(\hbar v_F \vec{k})^2 + (m + \eta \xi \Delta_{\textrm{SO}})^2},
\end{equation} 
where $ \eta $ and $ \xi $ are the eigenvalues of $ \tau_z $ and $ s_z $, features a spin and valley dependent energy gap. Since the spin in the $ z $-direction and the valley quantum number are conserved by $ \mathcal{H}_0 $, the Chern number    
	\begin{equation}
	\label{ChernNumberDefinition}
	n = \sum_{\substack{\textrm{$\alpha \in$ filled} \\ \textrm{bands}}} \frac{1}{2\pi} \int d^2 k \, \mathcal{F_{\alpha}},
	\end{equation}    
	with the Berry curvature $\mathcal{F}_{\alpha}=[\nabla_{\vec{k}}\times i \langle\phi_{\alpha\vec{k}}|\nabla_{\vec{k}}|\phi_{\alpha\vec{k}}\rangle]\cdot {\hat e}_{z}$ and where the spinor $\phi_{\alpha\vec{k}}$ is connected to the solution $\Lambda_{\alpha\vec{k}}(x)= \exp(i{\vec{k}}\cdot{\vec{x}})\phi_{\alpha\vec{k}}$\cite{HasanKaneTopologicalInsulators}, can be calculated for each spin and each valley separately \cite{Chan14},    
	\begin{equation}
	\label{eq:ChernNumber}
	n_{\eta \xi} = \frac{\eta}{2} \sign(m + \xi \eta \Delta_{\textrm{SO}}).
	\end{equation}    
	With these Chern numbers a topological $ \mathbb{Z}_2 $ invariant	
	\begin{equation}
	 \nu = \sum_{\eta} \frac{n_{\eta \uparrow} - n_{\eta \downarrow}}{2} \mod 2
	 = \left\{ \begin{aligned}
	 & 0 & \abs{m} > \Delta_{SO} \\
	 & 1 & \abs{m} < \Delta_{SO}
	 \end{aligned} \right.	 
	\end{equation} 	
	distinguishing the topologically trivial ($ \nu = 0 $) from the non trivial ($ \nu = 1 $) phase can be defined.	If the Chern numbers of the two half spaces differ, topological edge states exist at the interface due to the bulk-boundary correspondence \cite{HasanKaneTopologicalInsulators}. These edge states are valley chiral because the Chern number is proportional to $ \eta $, i.e.,  if a left-moving spin-up channel exists at the $ K $ valley, a right-moving spin-down channel will exist at the $ K' $ valley, which is a manifestation of $ \mathcal{H}_0 $ being time reversal invariant (see also Fig.~(\ref{fig-edgeStates})). Spin helicity (one Kramers pair of helical edge states) is achieved by tuning the electric fields to fulfill the condition
    \begin{equation}
	\label{eq:helicalCondition}
	\abs{m_i} < \Delta_\textrm{SO} < \abs{m_j},
	\end{equation} 
    $i,j=1,2$ ($i\neq j$). This is because in this regime the argument of the sgn-function $ m_{1, 2} \pm \Delta_{\textrm{SO}} $ changes (does not change) its sign when replacing $ m_1$ by $m_2$ for one of the two signs so that the channels do (do not) exist. Since the sign is given by $ \xi \eta $ one spin polarization is suppressed per valley and the resulting edge states are spin helical. This regime can for instance be accomplished by tuning the second electric field so that $ \Delta_{\textrm{SO}} < m_2 $ while the first one vanishes. Conversely, when tuning $ \vec{E}_{1} $ such that $ m_1 < - \Delta_{\textrm{SO}} $, edge states of both spin polarisations exist at the interface of the two half spaces and the edge states are spin degenerate. This crossover between spin helical and spin degenerate edge states can therefore be achieved by tuning the external electric fields. 
	
	By solving the Schr\"odinger equation $ \mathcal{H}_0(\vec{x}) \Lambda(\vec{x}) = \varepsilon\Lambda(\vec{x}) $ in both half spaces independently and matching their wave functions at the interface the dispersion relation of the edge states     
	\begin{equation}
	\varepsilon(k_x) = \pm\eta \hbar v_F k_x
	\label{eq:dispersion}
	\end{equation}      
	can be calculated without further approximations. It is independent of both the mass parameters and the spin orbit interaction strength. The prefactor $\pm$, however, depends on the concrete realisation of the mass parameters in both half spaces.	
	
    \section{Tunable Josephson effect}
        If the topological edge states in silicene are spin helical they effectively mimic those at the sample edge of a QSH insulator with the main difference being that the energy dispersions at zero energy do not cross at the $ \Gamma $ point in momentum space but are located at the $ K $ and $ K' $ Dirac points in the first Brillouin zone. When building a Josephson junction mediated by these edge states, i.e. upon including the proximity induced superconductivity described by $ \mathcal{H}_S $, where we choose the phase of the superconducting pairing potential of one of the superconductors to vanish while the other is equal to $ \phi $ (see Fig.~(\ref{fig-setup})), ABS localized inside the junction emerge \cite{FuKane2009JosephsonJunction}. We note in passing that electrically tunable Josephson junctions in silicene have been considered theoretically in the context of $0-\pi$- \cite{Wang14, Kuzmanovski16, Li16, Zhou17} and/or $\varphi_0$-junctions \cite{Wang16, Kuzmanovski16, Zhou17} where time-reversal symmetry is broken explicitly in these works.
        
We first develop an effective model by projecting the full Hamiltonian in the absence of $ \mathcal{H}_I $ onto the subspace spanned by the boundary states obtained above.

Without loss of generality we assume that the electric fields are arranged such that the channel on the $ K $ ($ K' $) valley is spin up (down) polarized (i.e. $ \abs{m_1} < \Delta_{SO} $ and $ m_2 < - \Delta_{SO} $). The field operators
\begin{equation}
\begin{aligned}
 c_{\uparrow}(x) &= \int dy f(y) \frac{1}{\sqrt{2}} \left[ c_{A K \uparrow}(\vec{x}) - c_{B K \uparrow}(\vec{x}) \right] , \\
 c_{\downarrow}(x) &= \int dy f(y) \frac{1}{\sqrt{2}} \left[ c_{A K' \downarrow}(\vec{x}) - c_{B K' \downarrow}(\vec{x}) \right]
\end{aligned}
\label{eq:fieldop}
\end{equation} 
with
\begin{equation}
 f(y) = \frac{1}{\sqrt{\mathcal{N}}} e^{- \frac{\abs{m + \Delta_\textrm{SO}}}{\hbar v_F} \abs{y}} ,
\quad
m = \left\{ \begin{aligned}
 & m_1 & y < 0 \\
 & m_2 & y > 0
\end{aligned} \right.
\end{equation} 
annihilate an electron in these edge states at position $ x $ along the channel direction and are connected to the annihilation operators of the edge eigenstates via a Fourier transformation in the $ x $ direction. The normalization factor $ \mathcal{N} $  is chosen such that $ \int dy \abs{f(y)}^2 = 1 $. The Hamiltonian projected (with projection operator ${\cal P}$) onto the subspace spanned by the valley chiral states can be written in the basis $ \Phi(x)=(c_{\uparrow}(x), c_{\downarrow}(x), c_{\downarrow}^\dagger(x), - c_{\uparrow}^\dagger(x))^T $ as
\begin{equation}
{\cal P}H{\cal P}=
\begin{pmatrix}
i\hbar v_F \partial_x & 0 & \Delta e^{-i\phi} & 0 \\
0 & -i\hbar v_F \partial_x & 0 & \Delta e^{-i\phi} \\
\Delta e^{i\phi} & 0 & -i\hbar v_F \partial_x & 0 \\
0 & \Delta e^{i\phi} & 0 & i\hbar v_F \partial_x \\
\end{pmatrix}.
\label{projectedH}
\end{equation}
\subsection{Short-junction limit}
\begin{figure}
            \includegraphics[width = 8cm]{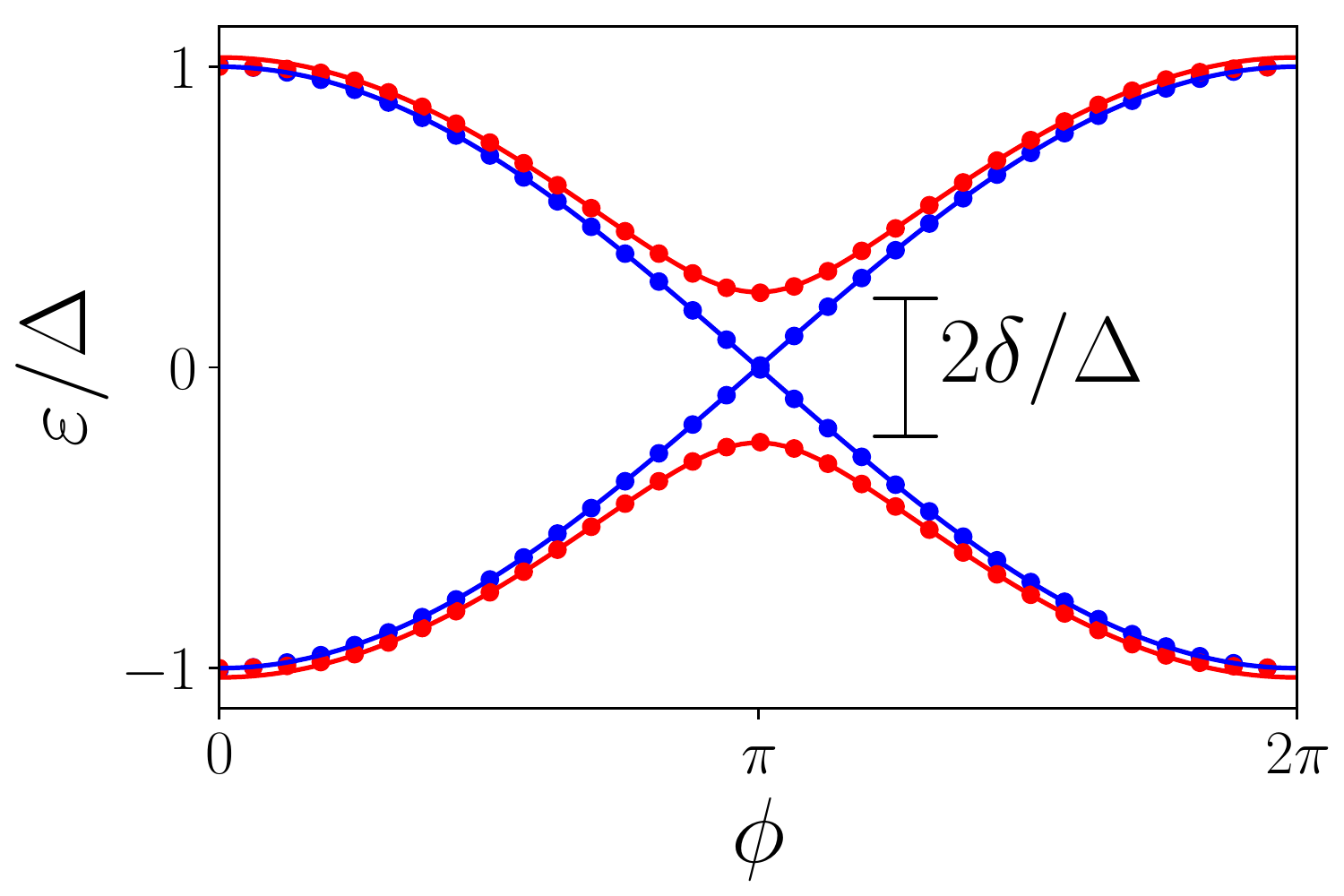}
            \caption{Energy-phase relation of the Andreev bound states. Without intervalley scattering (blue) the curves cross at zero energy and are $ 4 \pi $ periodic both in the helical and non-helical case. In the generic case, intervalley scattering $ \delta \rho_z \tau_x $ opens a gap at zero energy (red) in the non-helical regime and the energy-phase relation becomes $ 2 \pi $ periodic, whereas the crossing of Andreev bound states is protected by parity conservation in the topologically non-trivial regime with underlying helical edge states. The result obtained by implementing the low energy model on a square lattice of $ 50 \times 50 $ lattice sites (dotted lines) coincides well with the analytical prediction (solid lines).}
            \label{fig-results}
        \end{figure}

\begin{figure}
            \includegraphics[width = \columnwidth]{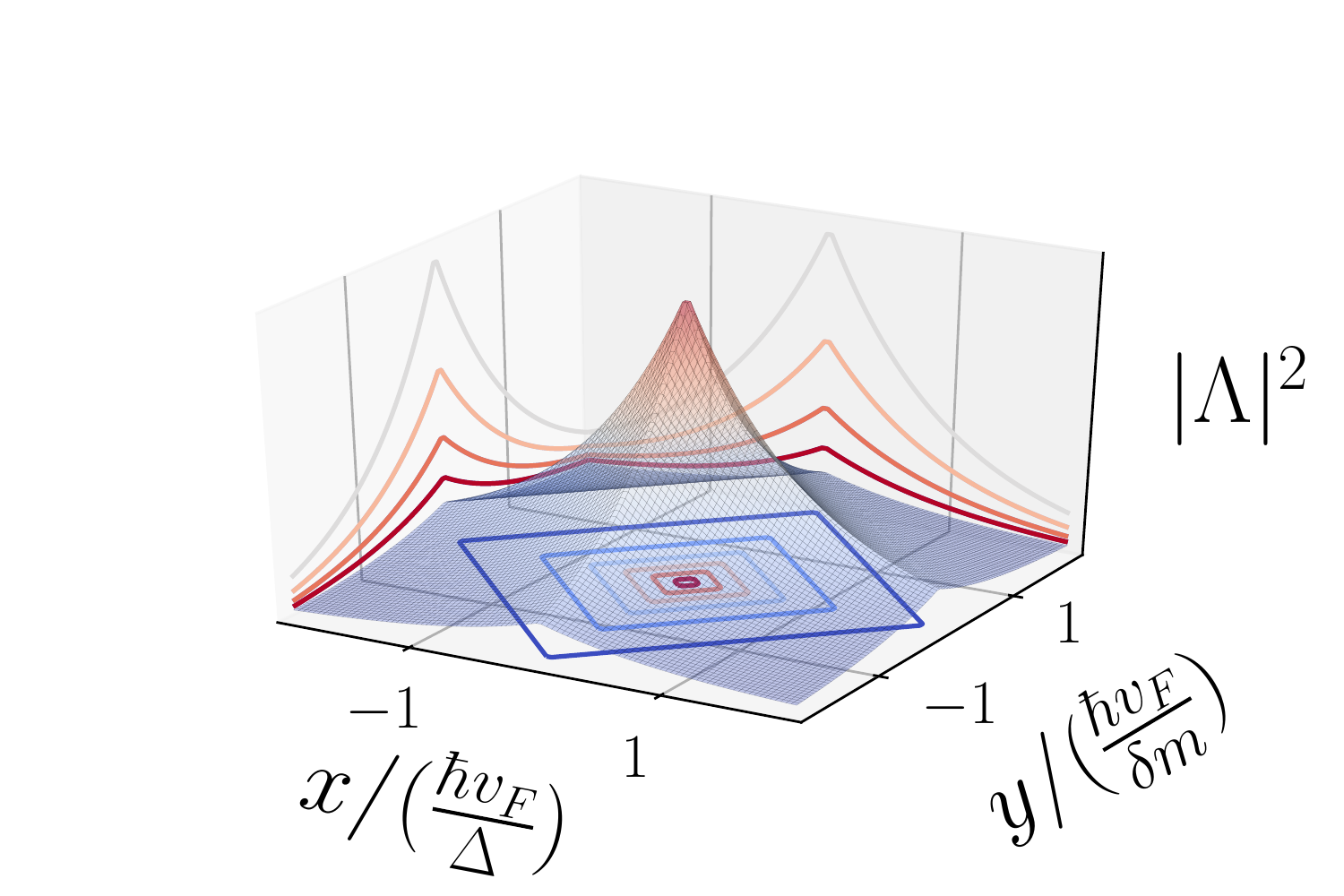}
            \caption{Probability density of the Majorana wave functions appearing for $\phi=\pi$. The degenerate wave functions are localized in $ x $-direction within the superconducting coherence length while the localization in $ y $-direction is dependent on the electric fields. The fields are chosen such that the mass parameters of both regions are symmetric around $ -\Delta_{SO} $ ($ m_{1, 2} = -\Delta_{SO} \pm \delta m $) so that the wave functions decay equally rapidly in the $ y $-direction on the length scale of $ \hbar v_F/\delta m $. The projections on the vertical planes are for specific values of $ x, y $.}
            \label{Majoranafigures}
        \end{figure}

Here, we assume that the two superconductors are located next to each other with their edges being perpendicular to the mass boundary, forming a short Josephson junction where $ l \ll \xi_0 $ with the superconducting coherence length $ \xi_0=\hbar v_F/\Delta $. Solving for the decaying wave-functions on both superconducting sides for $|\varepsilon|<\Delta$ and $ \mu = 0 $ and matching them at the junction boundary located at $x=0$ leads to two ABS      
        \begin{equation}
        \begin{aligned}
         \Gamma_1 &= \int dx \, \varphi(x) \frac{1}{\sqrt{2}} 
         \left[ e^{i\theta} c_{\uparrow}(x) + c_{\downarrow}^\dagger(x) \right]\\
         \Gamma_2 &= \int dx \, \varphi(x) \frac{i}{\sqrt{2}} 
         \left[ e^{-i\theta} c_{\downarrow}(x) - c_{\uparrow}^\dagger(x) \right] ,
        \end{aligned}
        \label{ABS}
        \end{equation}       
        where        
        \begin{equation}
        \begin{aligned}
         \varphi(x) &= \sqrt{\frac{\sqrt{\Delta^2 - \varepsilon^2}}{\hbar v_F}}
         e^{- \frac{\sqrt{\Delta^2 - \varepsilon^2}}{\hbar v_F} \abs{x}} \\
         \theta &= \arg \left( \varepsilon + i \sqrt{\Delta^2 - \varepsilon^2} \right)
         \end{aligned} .
        \end{equation}        
        They obey the energy-phase relation      
        \begin{equation}
        \label{eq:EnergyPhaseRelation}
         \varepsilon(\phi) = \pm \Delta \cos \left( \frac{\phi}{2} \right)
        \end{equation}         
        that is $ 4 \pi $-periodic (Fig.~\ref{fig-results}, blue), so that the Josephson current through the junction at zero temperature being proportional to $ \partial_{\phi} \varepsilon $ is also $ 4 \pi $-periodic. For $ \phi = \pi $, the ABS lie at zero energy and $ \theta \rightarrow \pi / 2 $, so that the linear combinations      
        \begin{equation}
         \gamma_1 = \left. \frac{(\Gamma_1 + \Gamma_2)}{\sqrt{2}} \right|_{\varepsilon = 0},
         \qquad
         \gamma_2 = \left. \frac{i(\Gamma_1 - \Gamma_2)}{\sqrt{2}} \right|_{\varepsilon = 0}
         \label{Majopairs}
        \end{equation}      
        are Majorana excitations due to the relation $ \Gamma_1^\dagger|_{\varepsilon = 0} = \Gamma_2|_{\varepsilon = 0} $. The corresponding Majorana wave functions are shown in Fig.~(\ref{Majoranafigures}) and are obtained from Eqs.~(\ref{ABS}) and (\ref{eq:fieldop}).
               
Tuning the electric fields such that the topological edge states are not spin helical but spin degenerate can be interpreted as introducing a second set of Kramers pair edge states with the opposite spin polarization. 

In the absence of intervalley scattering $ \mathcal{H}_I $, these two sets of Kramers pair edge states are independent of each other and both result in ABS with the $ 4 \pi $-periodic energy-phase relation \eqref{eq:EnergyPhaseRelation} which again translates into an overall $ 4 \pi $-periodic Josephson current. Since the intervalley scattering couples electrons with the same spin polarization of different Dirac cones, it induces scattering between the two sets of Kramers pair edge states in the absence of the superconductors or couples the two sets of ABS in the presence of the Josephson junction. Due to this coupling a gap of size $ 2 \delta $ opens at zero energy in the energy-phase relations of the ABS. The resulting energy-phase relation       
        \begin{equation}
         \varepsilon(\phi) = \pm \sqrt{ \Delta^2 \cos^2 \left( \frac{\phi}{2} \right) + \delta^2 }
        \end{equation}       
        is $ 2 \pi $-periodic (Fig.~\ref{fig-results}, red) which results in the Josephson current now also being $ 2 \pi $-periodic. 
        
 We compare the low-energy model with a numerical treatment of the full low-energy Hamiltonian Eq.~\eqref{eq:Hamiltonian} on a square lattice (see Appendix~\ref{AppA} for further details). We note that the opening of an energy gap is a consequence of the ABS not being protected in the topologically trivial phase, so that these results should be valid even for more general forms of intervalley scattering.
            
We note that the effect of intervalley scattering on the ABS in the spin-degenerate case depends on the chemical potential $\mu$, which, so far, we have set to zero. By repeating the calculations for the ABS without intervalley scattering and calculating the matrix elements between ABS of different valleys, the minimal energy of the ABS as a function of $\phi$ becomes $\delta/(1+(\mu/\Delta)^2)$. For small $\mu$ this corresponds  to a correction $-\delta(\mu/\Delta)^2$ that reduces the gap in the spectrum around $\phi=\pi$, however, it is parametrically small by the factor $(\mu/\Delta)^2$. In the opposite limit $\abs{\mu} \gg \Delta$, the effect of intervalley scattering is suppressed by the factor $(\Delta/\mu)^2$. This analysis shows that we can tune the influence of intervalley scattering on the ABS by the chemical potential $\mu$. 
     
Intravalley scattering can also be present, but has no effect on the ABS since it corresponds to forward scattering due to the valley chirality of all boundary states. In addition, we have neglected the much smaller Rashba effect compared to the intrinsic spin-orbit effect \cite{LowEnergyModelOfSilicene}. The Rashba spin orbit coupling could split the spin degeneracy in the spin degenerate regime, but would not lead to a change of the periodicity of the Josephson effect.

\subsection{Long-junction limit}    
\begin{figure}
	  \includegraphics[width = 8cm]{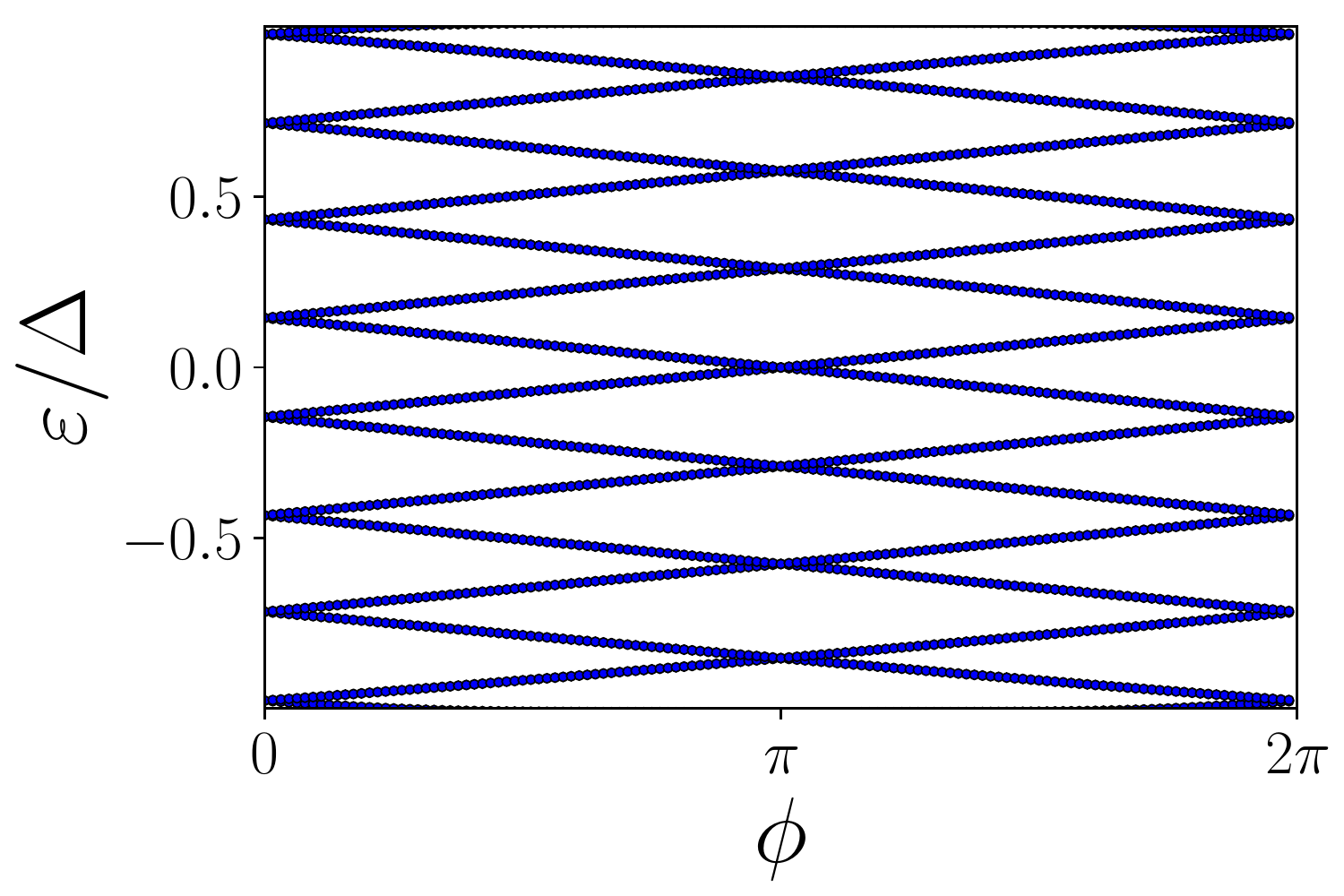}
	  \includegraphics[width = 7.5cm]{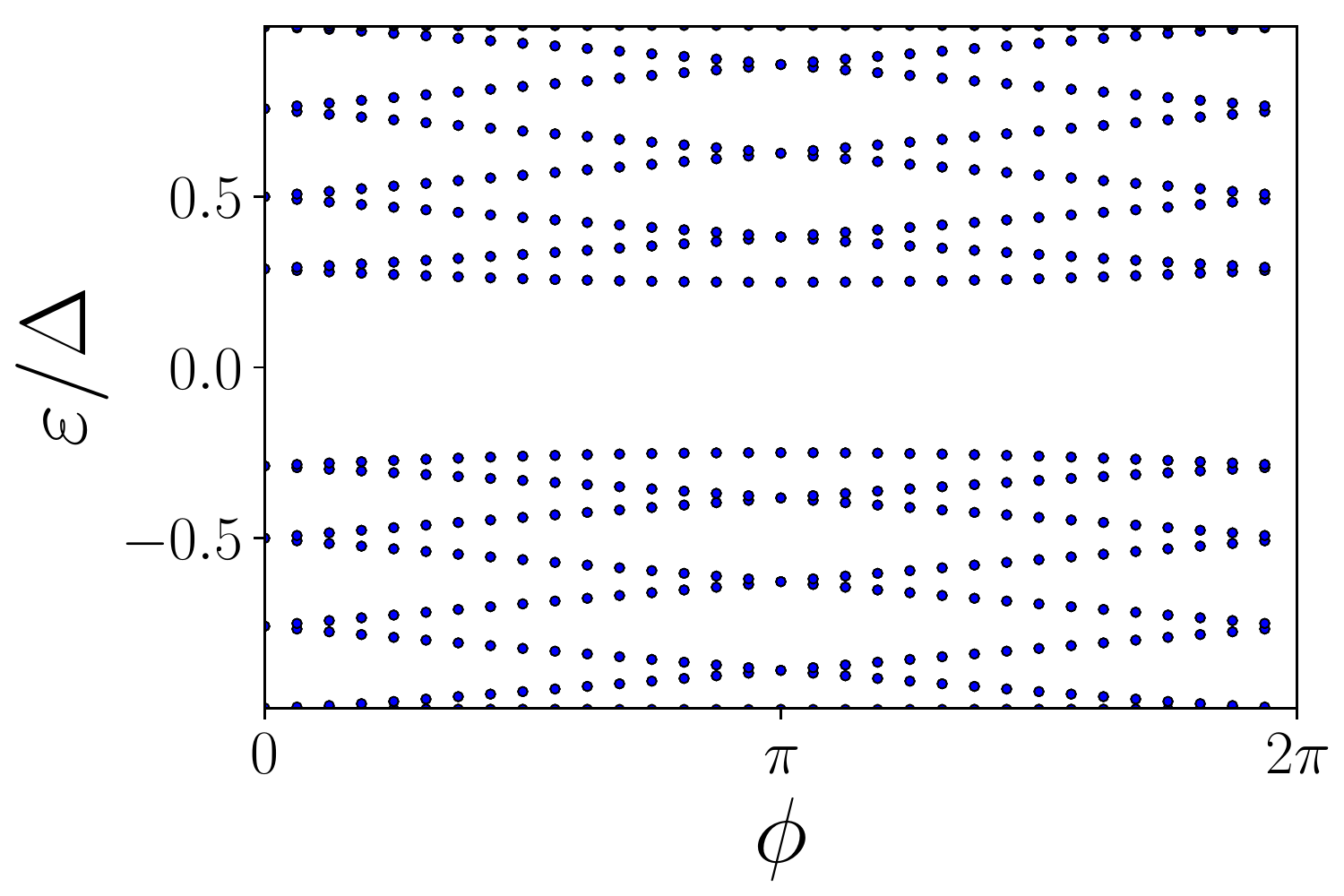}
	  \caption{Energy-phase relation of the Andreev bound states in the long-junction regime $ l \gg \xi_0 $ obtained numerically from the low energy model Eq.~(\ref{eq:Hamiltonian}) implemented on a square lattice of $ 100 \times 100 $ lattice sites in the helical regime. The length of the junction ($ 50 $ lattice sites) is large compared to the coherence length of the system ($ \sim 1.6 $ lattice sites) so that the junction is well in the long-junction regime. Top panel: No intervalley scattering, multiple ABS exist inside the junction and feature an energy-phase relation linear in $\phi$. Bottom panel: Andreev bound states with intervalley scattering induced by ${\cal H}_I$ where $ \delta = 0.25 \Delta $ in the spin-degenerate regime.}
	  \label{fig-longJunction}
      \end{figure}

  Our calculation of the ABS is valid in the short junction regime $l\ll \xi_0$. In a real experiment, however, the Josephson junction may be in the long-junction regime $l\gg \xi_0$ \cite{CriticalSupercurrentBeenakker}. With a Fermi velocity $ v_F = 5.42 \times 10^5 \, {\rm m/s} $ in silicene and a superconducting gap of the order of $ \Delta \sim 1\,{\rm meV}$, $\xi_0 \sim 357 \, {\rm nm} $ so that the junctions fabricated in Ref.~\citenum{Bocquillon2016} and \citenum{Deacon2017} for HgTe/CdTe quantum wells being $ 400 \, {\rm nm} $ and $ 600 \, {\rm nm} $ long, respectively would fall into the intermediate or long-junction regime. In the latter regime, the energy-phase relation has been theoretically shown to depend linearly on the phase difference $ \phi $ across the junction \cite{CriticalSupercurrentBeenakker, LinearJosephsonCurrentIshii, LinearJosephsonCurrentBardeen, LinearJosephsonCurrentSvidzinsky} while its periodicity stays unchanged. In Fig.~\ref{fig-longJunction}, we present results for the ABS energies using a numerical simulation of the low-energy model Eq.~\eqref{eq:Hamiltonian} implemented on a square lattice (see Appendix~\ref{AppA}). This and all other numerical tight-binding simulations were performed with the Kwant code \cite{Groth2014}.
      
 \section{Experimental realization} 
\begin{figure}
	  \includegraphics[width = \columnwidth]{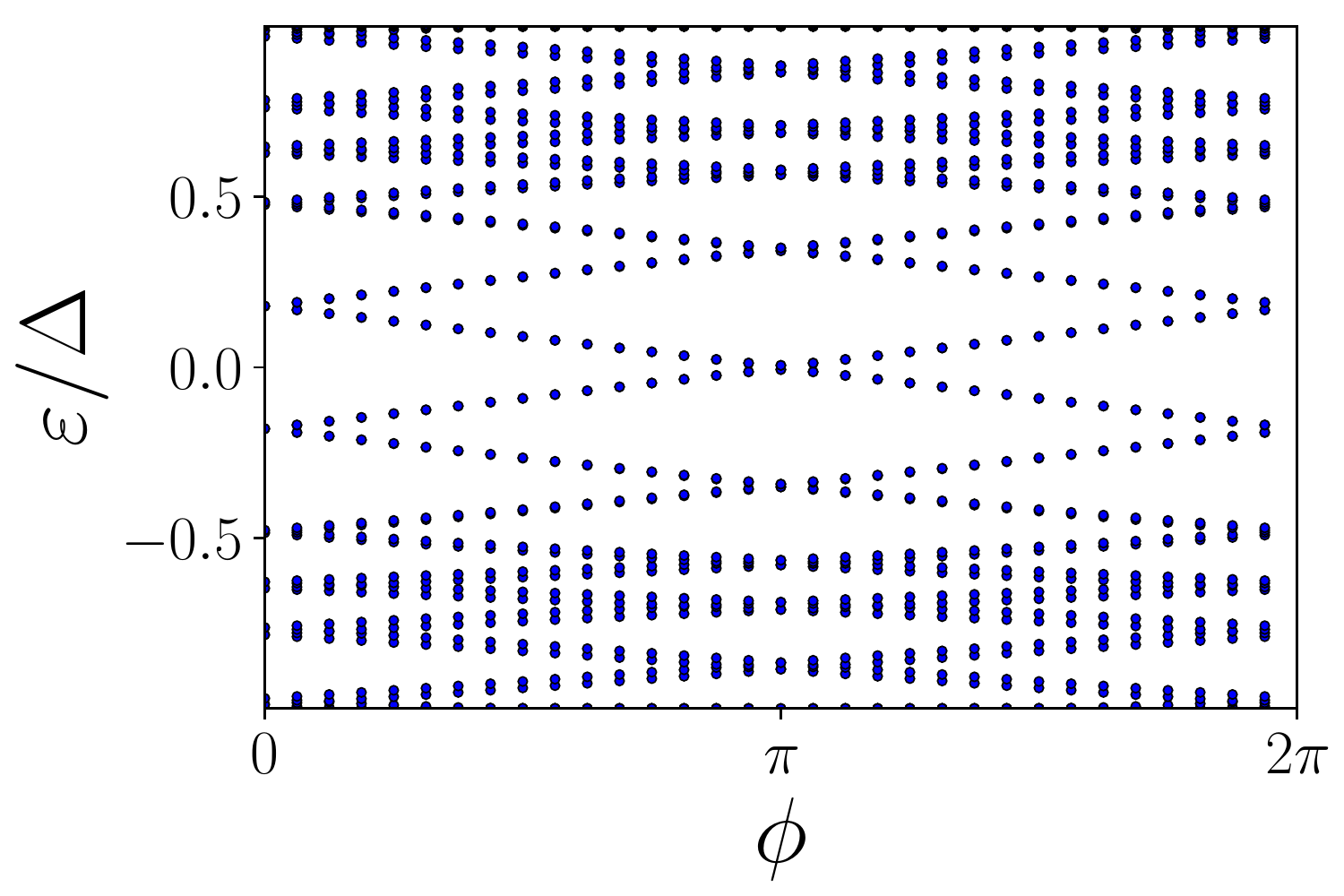}
	  \includegraphics[width = \columnwidth]{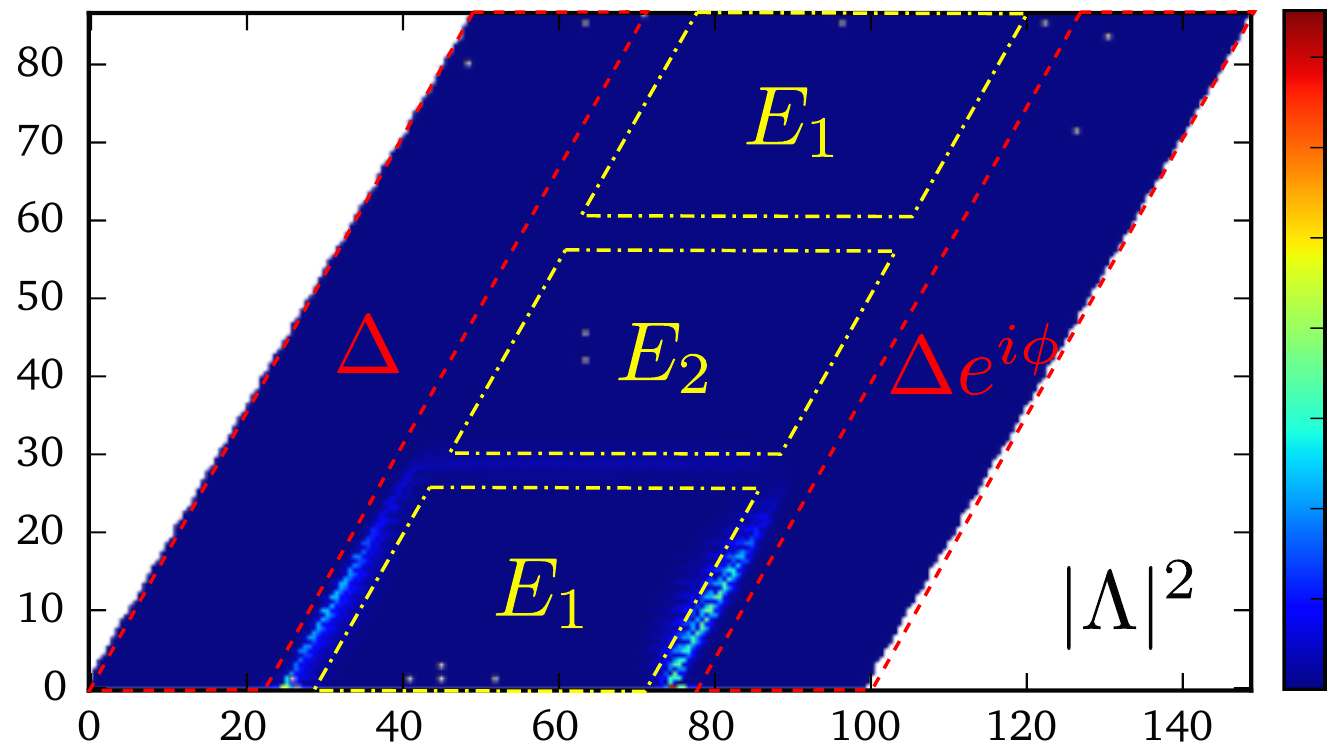}
	  \includegraphics[width = \columnwidth]{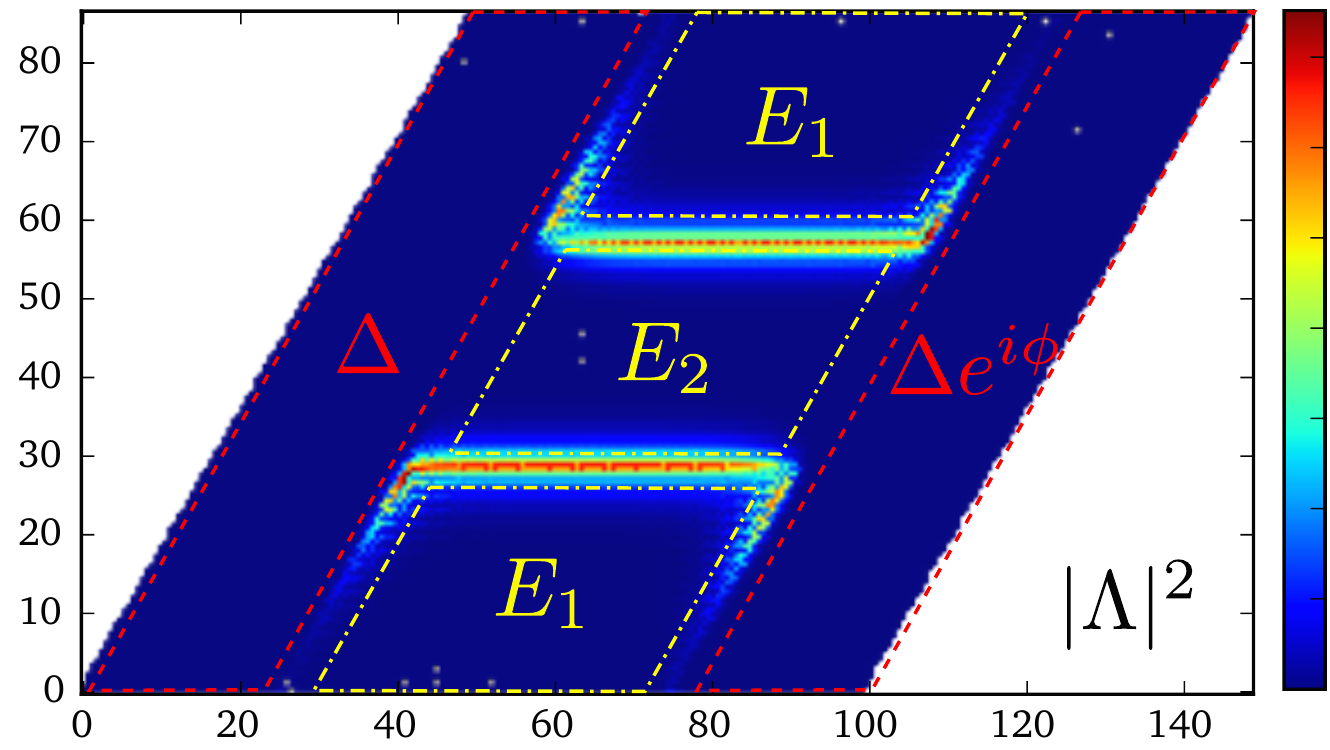}
	  \caption{Energy-phase relation (top panel) of in-gap states obtained by implementing a tight binding model of silicene on a hexagonal lattice of $ 100 \times 100 $ unit cells. The probability density (middle and bottom) are shown on a linear scale, where red colors correspond to high probability densities. The $ x- $ and $ y- $axes are given in units of the length of a single unit cell. The electric fields are only present inside the junction ($ 50 $ unit cells wide). The superconducting regions (red dashed boxes, middle and bottom panels) and regions with a finite electric field (yellow dashed boxes, middle and bottom panels) actually touch and are only drawn with a spatial gap to better visualize the probability density localized at the border between these regions. In addition to the energy-phase relations already present in the case for a finite electric field inside the superconducting regions (Fig.~\ref{fig-longJunction}) there exist in-gap states whose energies appear to be independent of the phase difference across the junction at $ \varepsilon \approx \pm\Delta/2 $. The probability density corresponding to such states (middle panel) is localized at the sides of the superconductors and not inside the junction like the ABS shown at the bottom for zero energy. Since the field $ E_1 $ tunes the region to be topologically trivial no edge channels are present at the top and bottom edges of the sample.}
	  \label{fig-noElectricalFieldInSC}
      \end{figure}

 The proposed setup could be implemented by using electric top and bottom gates to define the topologically confined channel. For instance, choosing $ m_1 = 0 $ and $ m_2 = -2 \Delta_{\textrm{SO}} $ (for which electric fields $ E_{1} = 0 $ and $ E_2 \approx - 2 \times 17 $ meV/\AA \, $ = -34 $ meV/\AA \, are needed \cite{Ezawa15, Drummond2012}) the spin helical edge states will lie inside an energy gap of $ \Delta_{\textrm{SO}} \approx 3.9 $ meV. Choosing $ m_1 = 2 \Delta_{\textrm{SO}} $ and $ m_2 = -2 \Delta_{\textrm{SO}} $ (for which electric fields $ E_{1,2} = \pm 2 \Delta_{\textrm{SO}} \approx \pm 34 $ meV/\AA \, are needed) spin degenerate edge states will again lie inside an energy gap of $ \Delta_{\textrm{SO}} \approx 3.9 $ meV. If the gates do not reach into the superconducting region the channel would not be present below the superconductors and we would assume that $m_1$ and $m_2$ are zero there.  We can treat this scenario numerically by describing the silicene sheet by a tight binding model with nearest neighbor hopping, inversion symmetry breaking staggered potentials, a Kane-Mele-type intrinsic spin orbit interaction \cite{Kane_2005} and s-wave superconductivity via the BdG-formalism (see Appendix~\ref{AppB}). We find that, in addition to the energy-phase relation of in-gap states already present with electric fields reaching into the superconducting regions (Fig.~\ref{fig-longJunction}, top panel), new in-gap states emerge (Fig.~\ref{fig-noElectricalFieldInSC}, top panel) at energies $ \varepsilon \approx \pm\Delta/2 $ which appear to be independent of the phase difference across the junction and are localized at the edges of the superconductors (Fig.~\ref{fig-noElectricalFieldInSC}, middle panel) in contrast to the ABS spread homogenously between the two superconductors along the topological channel (Fig.~\ref{fig-noElectricalFieldInSC}, bottom panel). These states are remnants of the topological edge states, which would propagate along the edges of the superconducting regions (but with vanishing $\Delta$) instead of straight into these regions, as would be the case for a finite electric field inside the superconducting regions (see Appendix~\ref{AppD}). However, as long as the electric fields merely vanish inside the superconducting regions, these states only occur at $ \varepsilon \approx \pm \frac{\Delta}{2} $ and leave the low energy excitations, and therefore the main aspects of our proposal, unchanged.
 In Appendix~\ref{AppC}, we calculate analytically these bound states located at the boundary between a superconductor and a region with a mass gap for silicene and show that for the parameters used in Fig. 6, an energy gap on the scale of $\Delta$ appears, similar to the numerics. The spectrum of these states is flat as a function of the superconducting phase difference $\phi$ since they are localized near one of the two superconductors with only an negligible overlap with the other superconductor
 \footnote{In the case, where the superconductors dope the silicene sheet below them, potential steps between the valley-chiral channels and superconducting regions develop. We numerically checked that the current phase relation is not influenced by such doping effects.}.       
        
 \section{Tunable poisoning in a 4$\pi$-junction}  
 \label{sec:poisoning}
\begin{figure}
	  \includegraphics[width = 8cm]{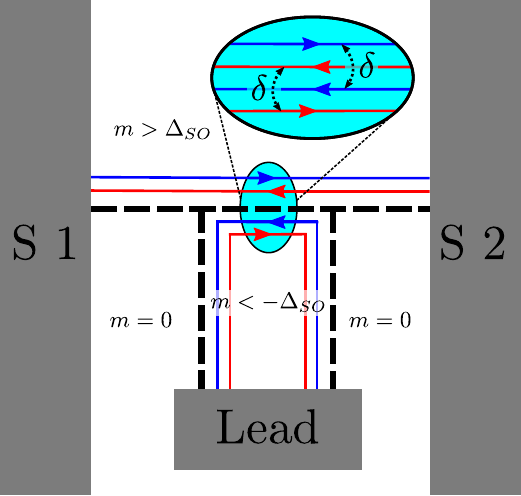}
	  \caption{Helical and valley chiral Josephson junction with tunable quasiparticle poisoning that can be used to switch between a $4\pi$- and $2\pi$-junction electrically. Blue (red) lines represent the spin up (down) polarized edge channels. Sections of spin helical and spin-degenerate channels can be controlled via tunable mass terms. The spin-degenerate regime can act---in the presence of intervalley scattering $\delta$---as a tunnel junction between the Andreev bound states formed in the helical and valley chiral Josephson junction and a second helical and valley chiral channel (of opposite spin) that is coupled to a lead allowing for dissipation. This lead has the same chemical potential as the two superconductors.}
	  \label{poisoning}
      \end{figure}

So far we have shown how one can electrically switch from a spin-helical regime to a spin-degenerate regime for the valley chiral channels, leading to generically different types of Josephson effects. Another promising direction is offered by utilizing the tunable mass-regions for an additional probe for the spin-helical channel, see Fig.~(\ref{poisoning}). The mass-terms, i.e. the gate voltages, can be tuned such that the helical channel is coupled to another helical channel---forming a small region of weakly coupled (by intervalley scattering) spin-degenerate valley-chiral states---that is further coupled dissipatively to an electron reservoir. Assuming a phase-biased Josephson junction formed via the spin-helical channel, the coupling to the dissipative channel, if present, can be used to relax the Josephson junction formed by the helical channel to its instantaneous ground state $|0(\phi)\rangle$, leading to a 2$\pi$-periodic Josephson current, despite the fact that the channel is helical. Removing the dissipative channel, by switching the electric gates, establishes the return to the 4$\pi$-periodic Josephson current.
 
To show this, we use the following effective model for the short junction regime (similar conclusions should also hold for the long-junction regime). The helical edge states in contact with the two superconductors having a phase difference of $\phi$ are represented by its instantaneous eigenstates via the Hamiltonian $H_0=(\varepsilon_a(\phi)\Gamma_a^{\dagger}\Gamma_a+\varepsilon_b(\phi)\Gamma_b^{\dagger}\Gamma_b)/2$. Using $\Gamma_a=\Gamma_b^{\dagger}$ and  $\varepsilon_a(\phi)=-\varepsilon_b(\phi)$ due to electron-hole symmetry, we rewrite $H_0=\varepsilon_a(\phi)(\Gamma_a^{\dagger}\Gamma_a -1/2)$. The many-particle states are consequently $|1(\phi)\rangle=\Gamma_a^{\dagger}|0(\phi)\rangle$ with energy $\varepsilon_a(\phi)/2$, and $|0(\phi)\rangle$ with energy $\varepsilon_0(\phi)=-\varepsilon_a(\phi)/2$. Also, it holds that $\Gamma_a|0(\phi)\rangle=0$. The operator $\Gamma_a$ is directly related to $\Gamma_{1,2}$ defined in Eq.~(\ref{ABS}) such that $\varepsilon_a(\phi)=|\Delta\cos(\phi/2)|$. We model the coupling of the spin-helical Josephson junction to the dissipative channel by a tunneling Hamiltonian $H_T=t\sum_\sigma c_{L\sigma}(0)c^{\dagger}_{R\sigma}(0)+{\rm h.c.}$, where we assume a pointlike tunneling region in the low-energy model. The tunneling matrix element $t$ originates from the microscopic form of intervalley scattering in the sample.

The field operators in the low-energy model are given for both sides $(L,R)$ by Eq.~(\ref{eq:fieldop}). We expand them on the side hosting the ABS ($L$) in the operators $\Gamma_1$ and  $\Gamma_2$ for a fixed phase difference $\phi$ and on the $R$-side in plane-wave states for the spin-helical channel $c_{R\sigma}(x)=(1/\sqrt{{\bar l}})\sum_{k}\exp(ikx)c_{Rk\sigma}$, see Appendix~\ref{App:pois}. Here, $c_{Rk\sigma}$ annihilates an electron with spin $\sigma$ and wave number $k$ in the dissipative helical channel and ${\bar l}$ is the quantization length for these channels. Without loss of generality we assume that  $\Gamma_1^{\dagger}$ creates a particle in the ABS with positive energy $\varepsilon(\phi)>0$. Then the tunneling rates $W_{\alpha\beta}$ that change the many-particle states of the Josephson junction from state $\alpha$ to state $\beta$ are given by Fermi's Golden rule rates (see Appendix~\ref{App:pois}). Here, $\alpha$, $\beta$ denote the two possible states of the junction $|0(\phi)\rangle$ and $|1(\phi)\rangle$. We derive the following results $W_{10}=\gamma_{t} \varphi^{2}(0) f(\varepsilon(\phi))$ and $W_{01}=\gamma_{t} \varphi^{2}(0)[1-f(\varepsilon(\phi))]$ where $\gamma_{t}$ is the normal state tunneling rate $2\pi\nu |t|^2/\hbar$ with $\nu$ the density of states per spin and length in the dissipative helical channel and $f(\varepsilon)=[1+\exp(\beta \varepsilon)]^{-1}$ is the Fermi function in the dissipative helical channel with $\beta=1/k_B T$ the inverse thermal energy. We assume that there is no voltage bias between the superconducting reservoirs and the normal conducting reservoir (lead) to which the dissipative channels are coupled.

The dynamical state of the Josephson junction is then described by the reduced density matrix with elements $\rho_{\alpha\beta}$. The off-diagonal elements $\rho_{12}=\rho_{21}^{*}$ decay exponentially with a rate given by $(W_{10}+W_{01})/2$, whereas the probabilities $\rho_{ii}$ have a finite stationary value given by $(\rho_{00}/\rho_{11})=(W_{01}/W_{10})$ with the constraint $\rho_{00}+\rho_{11}=1$ which gives the solution for the occupation probability of the excited state $|1(\phi)\rangle$, $\rho_{11}=W_{10}/(W_{10}+W_{01})=f(\varepsilon(\phi))$. For low temperatures $T\rightarrow 0$, we obtain the desired result $\rho_{11}\sim \exp[-\beta\varepsilon(\phi)]$ so that the junction stays in the ground state $|0(\phi)\rangle$. 

A phase-biased experiment would therefore exhibit a $2\pi$ Josephson effect when slowly sweeping $\phi$ in the presence of the dissipative channel. For such a switching experiment to work, the intrinsic poisoning rates of the Josephson junction (i.e. the poisoning rate in the absence of the dissipative channel), need to be slower than $\gamma_t$. We note that Ref.~\onlinecite{CriticalSupercurrentBeenakker} suggested a doubling of the critical current of a long current biased topological Josephson junction in the $4\pi$-regime (weak poisoning) compared to the $2\pi$-regime (strong poisoning). According to Ref.~\onlinecite{CriticalSupercurrentBeenakker}, the possibility to switch between the two regimes would require for our setup that $\gamma_{t}^{-1}\ll\tau_J\ll\tau_{\rm{qp}}$, where $\tau_J$ is the internal phase relaxation time of the Josephson junction and $\tau_{\rm{qp}}$ is the intrinsic quasiparticle poisoning time.

\section{Conclusions}
We have investigated in detail a novel type of tunable Josephson junction through a channel defined at a mass domain in silicene. Using the electrically tunable mass term in silicene in combination with the intrinsic spin-orbit coupling and s-wave superconductivity via the proximity effect, we have demonstrated the transition between a valley chiral and spin helical boundary state and a valley chiral and spin degenerate boundary state. In both regimes we have calculated the Andreev bound states in the short junction limit analytically and numerically in the long-junction regime using a discretization of the low-energy Dirac model as well as using the full tight-binding model on the honeycomb lattice using the KWANT code \cite{Groth2014}. In the valley chiral and spin helical regime, the Josephson effect resembles the one through a helical edge state of a two-dimensional topological insulator \cite{FuKane2009JosephsonJunction, CriticalSupercurrentBeenakker}. The Josephson current exhibits a fractional (4$\pi$) current phase relation in the absence of quasiparticle poisoning. In the valley chiral but spin-degenerate regime, the Josephson effect is generically of the usual $2\pi$ type as intervalley scattering will open a gap in the Andreev bound state spectrum around $\phi=\pi$. In the absence of intervalley scattering and in the presence of fermion parity conservation, the periodicity of the Josephson effect would depend on the global fermion parity, similar to the case of two separate spin helical edge states coupled by the same superconductors \cite{Crepin2014}. Intravalley scattering events and/or Rashba spin-orbit coupling will not change the periodicity of the Josephson current. We also investigated numerically experimentally feasible ways for implementing the tunable electric fields in the setup. 

We further proposed a scenario where a valley chiral and spin helical channel is subjected to tunable quasiparticle poisoning by coupling a part of the channel to another valley chiral and spin helical channel that is connected to a normal reservoir leading to dissipation. In that part, intervalley scattering, generically present, will at small temperatures cool the Josephson junction to its ground state. This quasiparticle poisoning is tunable electrically since the additional channel is only present if certain mass terms are present. This tunable Josephson effect would allow us to detect a topological Josephson junction exhibiting a $4\pi$ periodic current phase relation by comparing the critical current of the junction in the cases with and without the (tunable) quasiparticle poisoning. According to Ref.~\onlinecite{CriticalSupercurrentBeenakker}, the critical current of a long Josephson junction should differ by a factor of two in these two cases. This constitutes a new way to search for topological effects in Josephson junctions in silicene. 

Finally, we would like to emphasize that silicene is not the only material system with such desired tunable properties. In the recently proposed HgTe double layer quantum wells \cite{Michetti12} as well as in the topological insulator InAs/GaSb type II quantum wells \cite{Liu08,Qu15} such mass domains could be implemented with electric fields. To some extend also in single HgTe-based quantum wells, an electrically tunable coupling between helical edge states in constrictions has been predicted \cite{Krueckl11}. In both material systems signs of topological superconductivity in edge states have been seen experimentally \cite{Knez12, Wiedenmann2016, Deacon2017}. Another promising material system is bilayer graphene where an electrically tunable gap can be implemented by a voltage between the layers and valley chiral states appear at voltage domain walls \cite{Martin08} with the difference that there are two spin-degenerate states per valley. Such voltage induced mass-gaps in bilayer graphene have been experimentally realized in graphene transistors \cite{Oostinga08} and for the creation of gate-tunable quantum dots \cite{Allen12,Eich18}. The coupling of such a valley-chiral channel to a superconductor has been recently investigated in the context of Cooper pair splitting \cite{Schroer15}. The problem of a comparably small spin-orbit coupling in graphene was recently taken up by considering curved bilayer graphene \cite{Klinovaja12} or by combining bilayer graphene with transition metal dichalcogenides \cite{Gmitra15,Khoo17, Alsharari18}.
  
    \begin{acknowledgments}
      We thank Bj\"orn Trauzettel for useful feedback on the manuscript and the Research Training Group GrK1952/1 ``Metrology for Complex Nanosystems'',
        and the Braunschweig International Graduate School of Metrology B-IGSM for financial support.
    \end{acknowledgments}

    \begin{appendix}

    \section{Low energy tight binding model on a square lattice}
    \label{AppA}
        The tight binding approach of the low energy model on a square lattice
        can be obtained by first Fourier transforming \eqref{eq:Hamiltonian} to momentum (k)-space and then discretize real space on a square lattice with lattice constant $ b $. The spinor in momentum space is then represented as     
        \begin{equation}
         \psi_{\vec{k}} = \frac{1}{\sqrt{N}} \sum_{\vec{R}} e^{-i \vec{k} \cdot \vec{R}} \psi_{\vec{R}},
        \end{equation}      
    where $ \psi_{\vec{k}} $ is the Fourier transform of $ \Psi(\vec{x}) $ defined in Eq.~\eqref{eq:Basis}, $ N $ is the number of lattice sites and $ \vec{R} $ is the position vector of a lattice site of a square lattice. The wave vectors in the Hamiltonian $ k_i = \frac{1}{b} k_i b \approx \frac{1}{b} \sin(k_i b) $ can be approximated by a sine function. With the orthonormality relation   
    \begin{equation}
     \delta_{\vec{R}, 0} = \frac{1}{N} \sum_{\vec{k}} e^{i \vec{k} \cdot \vec{R}}
    \end{equation}    
    the kinetic part of the Hamiltonian transforms to  
    \begin{equation}
    \begin{aligned}
    &\sum_{\vec{k}} \psi_{\vec{k}}^\dagger [ \hbar v_F (k_x \rho_z \tau_z \sigma_x + k_y \rho_z \sigma_y) ] \psi_{\vec{k}} \\
    &\approx \frac{\hbar v_F}{b} \sum_{\vec{k}} \psi_{\vec{k}}^\dagger 
    \left( \sin(k_x b) \rho_z \tau_z \sigma_x + \sin(k_y b) \rho_z \sigma_y \right) \psi_{\vec{k}} \\
    &= - i \frac{\hbar v_F}{2b} \sum_{\vec{R}} \left( \psi_{\vec{R} + \vec{\delta}_1}^\dagger \rho_z \tau_z \sigma_x \psi_{\vec{R}}
    + \psi_{\vec{R} + \vec{\delta}_2}^\dagger \rho_z \sigma_y \psi_{\vec{R}} \right) + \textrm{H.c.}
    \end{aligned}
    \end{equation}    
    where   
    \begin{equation}
    \vec{\delta}_1 =
    \begin{pmatrix}
      b \\
      0
    \end{pmatrix},
    \qquad
    \vec{\delta}_2 =
    \begin{pmatrix}
      0 \\
      b
    \end{pmatrix} .
    \end{equation} 
    The spectrum $ \varepsilon = \pm \frac{\hbar v_F}{b} \sqrt{\sin(k_x b)^2 + \sin(k_y b)^2} $ of the particle and $ K $ valley subspace approximates the corresponding spectrum $ \varepsilon = \pm \hbar v_F \abs{\vec{k}} $ of the kinetic part of the low energy model \eqref{eq:Hamiltonian:Parts} for small wave vectors $ \vec{k} $ but deviates significantly from it for large wave vectors. Most notably, in contrast to the original low energy model \eqref{eq:Hamiltonian} the spectrum of the discretized model introduces three additional inequivalent Dirac cones at the corners of the first Brillouin zone. Since the topology of a system is linked intimately to the number of Dirac cones present in the spectrum, these additional Dirac cones would significantly alter results directly linked to topology, i.e. the number of topological edge states. We therefore need to include another term, which gaps the three spurious additional Dirac cones but leaves the Dirac cone centered at the $ \Gamma $ point unaltered. Furthermore this additional term must respect the symmetry of the system, that is time reversal symmetry. A possible term reads
    \begin{equation}
     \frac{\hbar v_F}{b} (2 - \cos(k_x b) - \cos(k_y b) ) \rho_z \sigma_z
    \end{equation} 
    which we can identify as the ``lattice generalization of the mass term'' introduced in Chap.~8.3 of Ref.~\citenum{Bernevig9780691151755}. For a more detailed discussion see also Ref.~\citenum{Zhou2016}. If the inversion symmetry breaking mass parameter $ m $ becomes finite, the now gapped additional Dirac cones can be closed again if $ m = -2 \frac{\hbar v_F}{b} $ or $ m = -4 \frac{\hbar v_F}{b} $. Furthermore, if $ m = - \frac{\hbar v_F}{b} $, then the resulting bands are non dispersive, so that numerical implementations should satisfy the constraint
    \begin{equation}
    \label{eq:Supplemental:Constaint}
     m > - \frac{\hbar v_F}{b}.
    \end{equation} 
    This additional term transforms as
    \begin{equation}
    \begin{aligned}
    \sum_{\vec{k}} &\psi_{\vec{k}}^\dagger \left[\frac{\hbar v_F}{b} (2 - \cos(k_x b) - \cos(k_y b) \right] \rho_z \sigma_z \psi_{\vec{k}} \\
    &= 2 \frac{\hbar v_F}{b} \sum_{\vec{R}} \left( \psi_{\vec{R}}^\dagger \rho_z \sigma_z \psi_{\vec{R}} \right) \\
    &\hspace{0.5cm}- \frac{\hbar v_F}{2b} \sum_{\vec{R}} \sum_{i = 1}^2 \left(  \psi_{\vec{R} + \vec{\delta}_i}^\dagger \rho_z \sigma_z \psi_{\vec{R}}
    + \textrm{H.c.} \right).
    \end{aligned}
    \end{equation} 
    Since all other terms in the low energy model Eq.~\eqref{eq:Hamiltonian} are momentum independent, their Fourier transform is trivial, so that the final tight binding model on a square lattice of the low energy model, that is the discretized low energy model, reads
    \begin{multline}
    \mathcal{H} = \sum_{i} \psi_{i}^\dagger \varepsilon \psi_{i}
    - t \sum_{\langle i, j \rangle} \psi_{i}^\dagger \rho_z \sigma_z \psi_{j} \\
    - i t \sum_{\langle i, j \rangle_x} \nu_{ij} \psi_{i}^\dagger \rho_z \tau_z \sigma_x \psi_{j}
    - i t \sum_{\langle i, j \rangle_y} \nu_{ij} \psi_{i}^\dagger \rho_z \sigma_y \psi_{j}
    \end{multline} 
    with
    \begin{equation}
    \begin{aligned}
    \varepsilon &= \left(m + 4t\right) \rho_z \sigma_z + \Delta_{\textrm{SO}} \rho_z s_z \tau_z \sigma_z + \mathcal{H}_{S} + \mathcal{H}_{I} , \\
    t &= \frac{\hbar v_F}{2b}
    \end{aligned} .
    \end{equation} 
    Here $ \langle i, j \rangle $ denotes all nearest neighbors and $ \langle i, j \rangle_i $ nearest neighbors in $ i $-direction. Furthermore $ \nu_{ij} = \pm 1 $ are a set of signs where $ \nu_{ij} = 1 $, if the site $ i $ is to the right or top of site $ j $.
    
    In Figs.~\ref{fig-results} and \ref{fig-longJunction} we report numerical tight binding simulations based on this discretized low energy model. In these simulations we set $ \frac{\hbar v_F}{b} = 1 $ as the characteristic energy scale of the system and measure all other energies in units of $ \frac{\hbar v_F}{b} $. The parameters were chosen as follows:
    \begin{center}
    \begin{tabular}{c p{1cm} c}
    spin helical regime && spin degenerate regime \\ \\
    \begin{tabular}{|c | c|}
    \hline
     $ \frac{\hbar v_F}{b} $ & $ 1 $ \\ \hline
     $ m_1 $ & $ 0.5 $ \\
     $ m_2 $ & $ 1 $ \\
     $ t $ & $ 0.5 $ \\
     $ \Delta_\textrm{SO} $ & $ 0.75 $ \\
     $ \Delta $ & $ 0.2 $ \\
     $ \delta $ & $ 0 $ \\ \hline
    \end{tabular}
    &&
    \begin{tabular}{|c | c|}
    \hline
     $ \frac{\hbar v_F}{b} $ & $ 1 $ \\ \hline
     $ m_1 $ & $ -0.5 $ \\
     $ m_2 $ & $ 1 $ \\
     $ t $ & $ 0.5 $ \\
     $ \Delta_\textrm{SO} $ & $ 0 $ \\
     $ \Delta $ & $ 0.2 $ \\
     $ \delta $ & $ 0.05 $ \\ \hline
    \end{tabular}
    \end{tabular}
    \end{center}
    so that in both cases the constraint Eq.~\eqref{eq:Supplemental:Constaint} is satisfied. While the hopping parameter $ t $ is always fixed in the chosen units and the two mass parameters $ m_i $ can be externally tuned by the applied electric fields, the last three parameters are material specific. Parameter values clearly illustrating their effects have been chosen. The spin orbit coupling strength for silicene is known and non-vanishing, unlike our assumption in the spin degenerate regime. However, the spin orbit coupling strength $ \Delta_\textrm{SO} $ only enters the Chern number Eq.~\eqref{eq:ChernNumber} in the argument of the sign-function and determines the localization length of the topological edge states, so only the spin orbit interaction strength relative to the induced mass parameters is of physical relevance here.

    \section{Tight binding model of the hexagonal lattice}
    \label{AppB}
        The microscopic tight binding model, to which the Hamiltonian \eqref{eq:Hamiltonian} is the low energy approximation, is given by a Kane-Mele-type Hamiltonian \cite{Kane_2005} with s-wave superconductivity formulated in the Bogoliubov de Gennes formalism on a honeycomb lattice     
	\begin{equation}
	\mathcal{H} = \sum_{i} c_{i}^\dagger \varepsilon_i c_{i}
	- t \sum_{\langle i, j \rangle} c_{i}^\dagger \rho_z c_{j} 
	+ i t_2  \sum_{\langle \langle i, j \rangle \rangle} 
	\nu_{ij} c_{i}^\dagger \rho_z s_z c_{j}
	\end{equation} 
	where $ c_{i} = (c_{i \uparrow}, c_{i \downarrow}, c_{i \downarrow}^\dagger, - c_{i \uparrow}^\dagger)^T $. Here the on-site energies of undoped silicene
	\begin{equation}
	 \varepsilon_i = \pm m_i \rho_z + \Delta_i (\cos(\phi) \rho_x + \sin(\phi) \rho_y)
	\end{equation}
	consist of a staggered potential and an s-wave superconducting pairing term and the plus (minus) sign applies to the $ A $ ($ B $) sublattice. Again, $ \langle i, j \rangle $ denotes all nearest neighbors and $ \langle \langle i, j \rangle \rangle $ all next nearest neighbors. The Haldane phases $ \nu_{ij} = \pm 1 $ are a set of signs depending on the two nearest neighbor bonds connecting the next nearest neighbors $ i $ and $ j $. If we have to take a left (right) turn when moving from site $ j $ to $ i $ the Haldane phase equals $ +1 $ ($ -1 $). If we take the hopping energies to be
	\begin{equation}
	 t = \frac{2}{3} \frac{\hbar v_F}{a}, \qquad
	 t_2 = \frac{\Delta_\textrm{SO}}{3\sqrt{3}}
	\end{equation} 
	where $ a $ is the distance between two nearest neighbors, this tight binding Hamiltonian results in the low energy model Eq.~\eqref{eq:Hamiltonian} at low energies. The characteristic energy scale of this model is the nearest neighbor hopping energy $ t $. In the numerical results presented in Fig.~\ref{fig-noElectricalFieldInSC} we set this characteristic energy to unity and measured all energies in units of $ t $. The parameters were chosen as follows
    \begin{center}
    \begin{tabular}{|c | c|}
    \hline
     $ t $ & $ 1 $ \\ \hline
     $ m_2 $ & $ 0.5 $ \\
     $ m_1 $ & $ 1 $ \\
     $ \Delta_\textrm{SO} $ & $ 0.75 $ \\
     $ \Delta $ & $ 0.2 $ \\ \hline
    \end{tabular}
    \end{center}
    where again the values for the mass parameters $ m_{1,2} $ can be varied externally and the absolute value of the superconducting paring potential is induced externally by the proximity effect. The physical strength of the spin orbit interaction may differ from the assumed value $ \Delta_\textrm{SO} = 0.75 t $, however, the number of existing edge channels again only depends on the spin orbit strength compared to the mass parameters.

\section{Electric fields inside superconducting regions}
\label{AppD}

\begin{figure}
\includegraphics[width = \columnwidth]{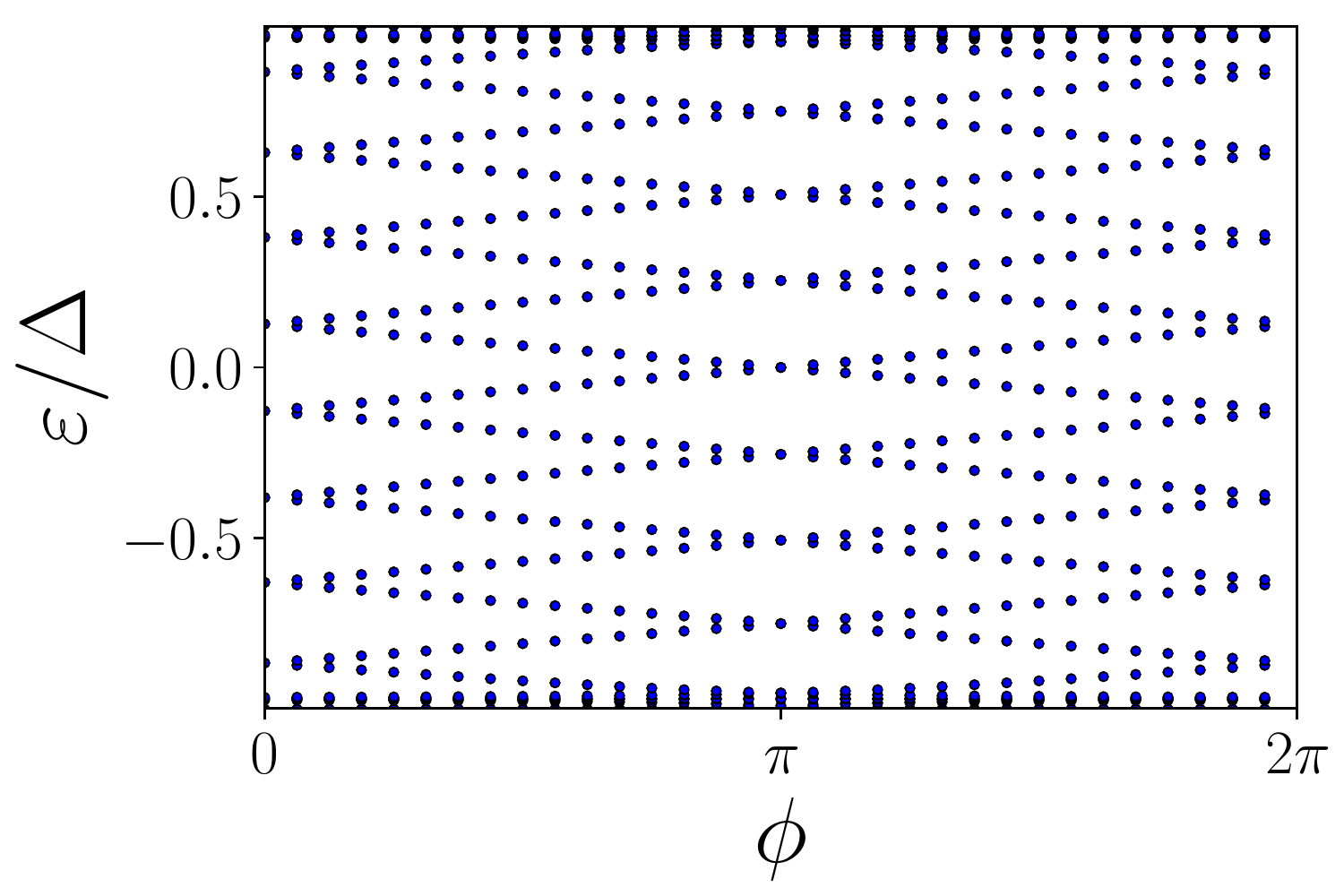}
\includegraphics[width = \columnwidth]{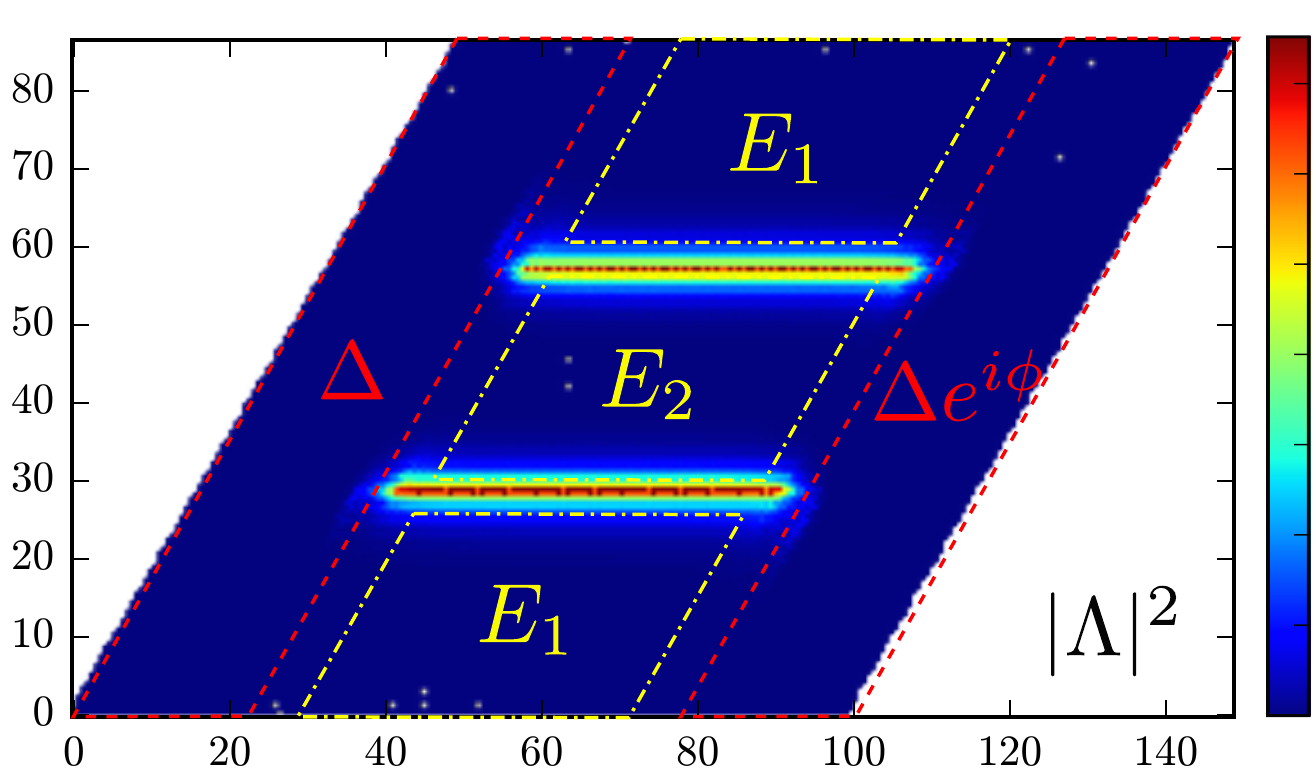}
\caption{Energy-phase relation (top) for the same setup as in Fig.~\ref{fig-noElectricalFieldInSC} with the only difference being that the electric fields now reach $ 5 $ lattice sites into the superconducting regions. Unlike the corresponding probability density of Fig.~\ref{fig-noElectricalFieldInSC} (bottom) the wave function at zero energy (bottom) now is well localized in the $ y $ direction also at the boundary of the superconductors.}
\label{fig-electricalFieldInSmallSCRegion}
\end{figure}

To better understand the existence of the additional in gap states at energies $ \varepsilon \approx \pm \Delta/2 $ existing in the case for vanishing electric fields inside the superconducting regions we also calculate the energy-phase relation for the same parameters as in Fig.~\ref{fig-noElectricalFieldInSC} with the only difference that the electric fields can now reach $ 5 $ lattice sites into the superconducting regions (Fig.~\ref{fig-electricalFieldInSmallSCRegion}, top). In this case the additional dispersionless states from Fig.~\ref{fig-noElectricalFieldInSC} are no longer present. Furthermore, when comparing the probability densities of a zero energy state ((Fig.~\ref{fig-noElectricalFieldInSC}, bottom) and (Fig.~\ref{fig-electricalFieldInSmallSCRegion}, bottom)) we note, that the bound state is well localized in the $ y $ direction even at the boundaries to the superconducting regions if the electric fields reach into the superconducting regions. Both observations support the premise that the additional states at energies $ \varepsilon \approx \pm \Delta/2 $ are remnants of the topological edge states without the superconducting pairing, because if the electric fields reach into the superconducting regions, the topological edge states are localized inside the superconducting regions and are thus suppressed by the superconducting pairing potential. If, on the other hand, the electric fields like in Fig.~\ref{fig-noElectricalFieldInSC} do not reach into the superconducting regions, the topological edge states are localized at the boundary of the superconducting regions and thus states localized at the boundary between the superconductor and the regions with a finite electric field can form. Furthermore, we note that the additional localization in the $ y $ direction at the boundary of the superconducting regions does not alter the energy-phase relation.
    
\section{Boundary states perpendicular to the channel}
\label{AppC}
We provide analytical support to our above argument for the existence of the in-gap states, 
which are independent of the superconducting phase difference, shown in Fig.~\ref{fig-noElectricalFieldInSC}. 
Specifically, we show that the states are subgap bound states at 
the interface between the normal region with a finite electric field and the superconducting region 
with a vanishing electric field, and hence do not carry supercurrent between two superconductors.  
To this end, we calculate the dispersion relation for in-gap states at the interface 
by solving the following BdG equation: 
\begin{gather}
\mathcal{H}(\vec{x}) \Psi(\vec{x}) = \varepsilon \Psi(\vec{x}), \\
\mathcal{H} = 
\begin{pmatrix}
\mathcal{H}_0(\vec{x}) & \Delta(\vec{x}) \\
\Delta^{*}(\vec{x}) &  -\mathcal{H}_0(\vec{x})
\end{pmatrix},
\end{gather}
where $\Psi(\vec{x})$ is the wave function in the Nambu basis, as defined in Eq.~\eqref{eq:Basis}, and 
$\mathcal{H}_0(\vec{x})$ is the Hamiltonian for silicene including a staggered potential, 
\begin{align}
\mathcal{H}_0(\vec{x}) =v_F p_x \sigma_x \tau_z + v_F p_y \sigma_y + \Delta_{\text{SO}} \sigma_z \tau_z s_z + m(\vec{x})\sigma_z.  
\end{align}
The matrices $\sigma_i$, $\tau_i$, $s_i$ are Pauli matrices acting on the spaces of sublattice, valley, and spin, respectively. 
The position-dependent variables $m(\vec{x})$ and $\Delta(\vec{x})$ are given by 
\begin{align}
m(\vec{x})=
\begin{cases}
m & \text{for $x > 0$}\\
0 & \text{for $x \leq 0$}
\end{cases}, \\
\Delta(\vec{x}) = 
\begin{cases}
0 & \text{for $x \geq 0$}\\
\Delta & \text{for $x < 0$}
\end{cases}, 
\end{align}
and they are uniform along the $y$-direction. Here we assume that chemical potential is zero. 
As the spin and valley degrees of freedom are conserved quantities in this model, 
we can reduce $\mathcal{H}(\vec{x})$ to a $4 \times 4$ matrix as 
\begin{align}
\mathcal{H}'(\vec{x}) = 
\begin{pmatrix}
\mathcal{H}'_0(\vec{x}) & \Delta(\vec{x}) \\
\Delta^{*}(\vec{x}) &  -\mathcal{H}'_0(\vec{x})
\end{pmatrix},
\end{align}
where
\begin{align}
\mathcal{H}'_0(\vec{x}) = \eta v_F p_x \sigma_x + v_F p_y \sigma_y + \eta \xi\Delta_{\text{SO}} \sigma_z  + m(\vec{x})\sigma_z.
\end{align}
Here, $\eta= \pm 1$ and $\xi= \pm 1$ are eigenvalues of $\tau_z$ and $s_z$, respectively. 
Below, we solve $\mathcal{H}'(\vec{x})$ in each of regions, $x<0$ and $x>0$.

In the superconducting region of $x < 0$, the reduced BdG equation is given by 
\begin{align}
\begin{pmatrix}
\mathcal{H}'^{S}_0  & \Delta \\
\Delta^{*} &  -\mathcal{H}'^{S}_0
\end{pmatrix} 
\begin{pmatrix}
\Psi^{e}_{S}(\vec{x}) \\
\Psi^{h}_{S}(\vec{x})
\end{pmatrix}
=\varepsilon 
\begin{pmatrix}
\Psi^{e}_{S}(\vec{x}) \\
\Psi^{h}_{S}(\vec{x})
\end{pmatrix},\label{S_eqn}
\end{align} 
with $\mathcal{H}'^{S}_0 = \eta v_F p_x \sigma_x + v_F p_y \sigma_y + \eta \xi\Delta_{\text{SO}} \sigma_z$. 
The bulk dispersion relation in this region is given by 
\begin{align}
\varepsilon = \pm \sqrt{\hbar^2 v^2_{F} k^{2}_x + \hbar^2 v^2_{F} k^{2}_y + \Delta^{2}_{\text{SO}}+|\Delta|^2}. 
\end{align}
Then a complex wave vector within the energy gap in the $x$ direction is 
\begin{align}
k_x &= -i \kappa_{S} \nonumber\\
&= -i (\hbar v_F)^{-1} \sqrt{\hbar^2 v^2_{F} k^{2}_y + \Delta^{2}_{\text{SO}}+|\Delta|^2 - \varepsilon^2}.\label{S_momentum}
\end{align} 
Note that we only consider $k_x = -i \kappa_{S}$ with $\kappa_{S} > 0$, as an imaginary momentum 
$k_x = i \kappa_{S}$ corresponds to a wave function which diverges as $x \rightarrow -\infty$.  
From Eqs.~\eqref{S_eqn} and \eqref{S_momentum}, we have 
\begin{align}
\mathcal{M} \Psi^{e}_{S}(\vec{x}) + \Delta \Psi^{h}_{S}(\vec{x}) = 0, \label{S_ftn}
\end{align}
where $\mathcal{M}$ is the $2 \times 2$ matrix given by
\begin{align}
\mathcal{M}=-i\eta v_F \hbar \kappa_{S} \sigma_x + v_F \hbar k_y \sigma_y + \eta \xi\Delta_{\text{SO}} \sigma_z - \varepsilon.
\end{align}
Eq.~\eqref{S_ftn} is used below in Eq.~\eqref{BMatching} to obtain a dispersion relation at the interface. 

Next we calculate wave functions in the normal region by solving the following equation, 
\begin{align}
\begin{pmatrix}
\mathcal{H}'^{N}_0 & 0 \\
0 &  -\mathcal{H}'^{N}_0
\end{pmatrix} 
\begin{pmatrix}
\Psi^{e}_{N}(\vec{x}) \\
\Psi^{h}_{N}(\vec{x})
\end{pmatrix}
=\varepsilon 
\begin{pmatrix}
\Psi^{e}_{N}(\vec{x}) \\
\Psi^{h}_{N}(\vec{x})
\end{pmatrix},\label{N_eqn}
\end{align}
where
$\mathcal{H}'^{N}_0 = \eta v_F p_x \sigma_x + v_F p_y \sigma_y + \eta \xi\Delta_{\text{SO}} \sigma_z  + m\sigma_z$.
The bulk dispersion relation for both an electron and a hole in the normal region is given by 
\begin{align}
\varepsilon = \pm \sqrt{\hbar^2 v^2_{F} k^{2}_x + \hbar^2 v^2_{F} k^{2}_y + (\eta \xi \Delta_{\text{SO}}+ m)^2},
\end{align}
and a complex wave vector in the $x$ direction corresponding to a wave function decaying as $x \rightarrow \infty$ is 
\begin{align}
k_x &= i \kappa_{N} \nonumber\\
&= i (\hbar v_F)^{-1} \sqrt{\hbar^2 v^2_{F} k^{2}_y + (\eta \xi \Delta_{\text{SO}}+ m)^2 - \varepsilon^2}.
\end{align}
The electron and hole components of the wave function are given by 
\begin{align}
\Psi^{e}_{N}(\vec{x}) &= c_e e^{-\kappa_N x + i k_y y} 
\begin{pmatrix}
\hbar v_F (i \eta \kappa_N - i k_y)\nonumber\\
\varepsilon - (\eta \xi \Delta_{\text{SO}}+ m)
\end{pmatrix}, \\
\Psi^{h}_{N}(\vec{x})&= c_h e^{-\kappa_N x + i k_y y} 
\begin{pmatrix}
\hbar v_F (i \eta \kappa_N - i k_y)\\
-\varepsilon - (\eta \xi \Delta_{\text{SO}}+ m)
\end{pmatrix}, \label{N_ftn}
\end{align}
where $c_e$ and $c_h$ are coefficients. 
By matching the wave functions, $\Psi^{e/h}_{S}(\vec{x}) = \Psi^{e/h}_{S}(x,y)$ given in Eq.~\eqref{S_ftn} 
and $\Psi^{e/h}_{N}(\vec{x})=\Psi^{e/h}_{N}(x,y)$ in Eq.~\eqref{N_ftn}, at $x=0$, 
\begin{align}
\begin{pmatrix}
\Psi^{e}_{S}(0,y) \\
\Psi^{h}_{S}(0,y)
\end{pmatrix}=
\begin{pmatrix}
\Psi^{e}_{N}(0,y) \\
\Psi^{h}_{N}(0,y)
\end{pmatrix}, \label{BMatching}
\end{align} 
we have
\begin{align}
c_e \mathcal{M}
\begin{pmatrix}
i \hbar v_F (\eta \kappa_N - k_y)\nonumber\\
\varepsilon - (\eta \xi \Delta_{\text{SO}}+ m)
\end{pmatrix} 
= c_h \Delta  
\begin{pmatrix}
-i \hbar v_F (\eta \kappa_N - k_y)\\
\varepsilon (\eta \xi \Delta_{\text{SO}}+ m)
\end{pmatrix}, \nonumber
\end{align}
from which we derive the dispersion relation at the interface. 
After some algebra, we find the simplified form as 
\begin{align}
\varepsilon^2 - \Delta^2_{\text{SO}}-\eta \xi\Delta_{\text{SO}} - \hbar^2 v^2_{F} k^{2}_y &= \hbar^2 v^2_{F} \kappa_S \kappa_N. 
\end{align}
This equation has a real solution $\varepsilon$ given by 
\begin{align}
\varepsilon = \pm \sqrt{\hbar^2 v^2_{F} k^{2}_y + |\Delta|^2 \frac{(\eta \xi \Delta_{\text{SO}}+ m)^2}{m^2+|\Delta|^2}},
\end{align}
provided that 
\begin{align}
Z = \left(1+ \frac{\eta \xi \Delta_{\text{SO}}}{m} \right) \left(1-\frac{\eta \xi \Delta_{\text{SO}} m }{|\Delta|^2} \right) >0. \label{Z_term}
\end{align}

We examine the result with two sets of parameters used in Fig.~\ref{fig-noElectricalFieldInSC}. 

{\it Case 1}. $m=1, \Delta_{\text{SO}}=0.75$, and $\Delta = 0.2$. 
In this case, $Z > 0$ is satisfied when $\eta \xi = -1$, and the solution is given by 
\begin{align}
\varepsilon(k_y) = \pm \sqrt{\hbar^2 v^2_{F} k^{2}_y + |\Delta|^2 \frac{(\Delta_{\text{SO}} - m)^2}{m^2+|\Delta|^2}},
\end{align} 
which has a gap at $k_y = 0$ of size $\varepsilon(0) \approx \pm 0.245 |\Delta|$. 

{\it Case 2}. $m=0.5, \Delta_{\text{SO}}=0.75$, and $\Delta = 0.2$. 
In this case, both cases of $\eta \xi = 1$ and $-1$ give $Z < 0$, and hence no solution exists within the gap.

 \section{Poisoning rates $W_{10}$ and $W_{01}$ }
 \label{App:pois}
Here, we present the details of the calculation for the poisoning rates and the stationary solution for the ABS occupation presented in subsection \ref{sec:poisoning}. 

We start with the solution for the ABSs
given in Eqs.~(\ref{ABS}). Fixing the superconducting phase difference $\phi$, we have two solutions of opposite energy $\Gamma_1(\varepsilon(\phi))\equiv \Gamma_1(\phi)$ and $\Gamma_2(-\varepsilon(\phi))\equiv \Gamma_2(\phi)=\Gamma_1^{\dagger}(\phi)$, due to particle-hole symmetry. Without loss of generality, we assume $\varepsilon(\phi)>0$.

A complete set of states are the eigenstates of the Hamiltonian $H$ (cf. Eq.~(\ref{eq:Hamiltonian})) with operators $\{\gamma_n\}$. Projecting $H$ to the low energy space spanned by the valley-chiral edge states, the field operator for these edge states is given by $\Phi(x)=(c_{\uparrow}(x),c_{\downarrow}(x),c_{\downarrow}^{\dagger}(x),-c_{\uparrow}^{\dagger}(x))^{T}$ defined above Eq.~(\ref{projectedH}). Near zero energy, and close to $\phi=\pi$, the low energy subspace is spanned by the ABSs with energies $|\varepsilon| < \Delta$ (cf. Eqn.~(\ref{ABS})), which we write here in the basis of $\Phi(x)$ as $\Gamma_1(\phi)=\int dx \varphi(x)(1/\sqrt{2})\left[\exp(i\theta(\varepsilon)),0,1,0\right]^T\Phi(x)$ (and $\Gamma_2(\phi)=\Gamma_1^{\dagger}(\phi)$). We model the coupling to the other two channels of opposite spin-valley chirality which are present at the same spatial position, but only along a finite region (see Fig.~(\ref{poisoning}),  by a tunneling Hamiltonian $H_T=t\sum_\sigma c_{L\sigma}(0)c^{\dagger}_{R\sigma}(0)+{\rm h.c.}$. Here, we consider a pointlike tunneling region in the low-energy model. The tunneling matrix element $t$ represents the microscopic form of intervalley scattering in the sample. Let $L$ stand for the channels coupled to the superconductors with ABS $\Gamma_1$ and  $\Gamma_2$, whereas $R$ labels the channels that are coupled to the dissipative reservoir (denoted as "Lead" in Fig.~(\ref{poisoning})). We expand $\Phi_{L}(x)$ in the low-energy ABSs as $\Phi_{L}(x)\approx (\varphi(x)/\sqrt{2})[\exp[-i\theta(\varepsilon(\phi))]\Gamma_1,\Gamma_1^{\dagger},\Gamma_1,-\exp[i\theta(\varepsilon(\phi))]\Gamma_{1}^{\dagger}]^{T}$. The field operators of the $R$-side we expand in the plane-wave states (wave-number $k$, spin $\sigma$) of the spin-helical channels $c_{R\sigma}(x)=(1/\sqrt{{\bar l}})\sum_{k}\exp(ikx)c_{Rk\sigma}$, where ${\bar l}$ is the quantization length for these channels. 

In the remainder of this appendix, we present the details of the calculation of the rates $W_{\alpha\beta}$ that switch the Josephson junction from a many-particle state $\alpha$ to a state $\beta$ by resorting to a Fermi's Golden Rule approach. We first rewrite $H_T$ in the appropriate excitations as 
\begin{multline}
\label{pois:app:eq0}
H_T=(t^*\varphi(0)/\sqrt{2{\bar l}}) \\
\times\sum_k\left(c_{Rk\uparrow}\Gamma_1^{\dagger}(\phi)e^{i\theta(\varepsilon(\phi))}+c_{Rk\downarrow}\Gamma_{1}(\phi)\right) 
+{\rm h.c.}
\end{multline}
The rate from the ground state $|0(\phi)\rangle_{\rm {ABS}}$ of the Josephson junction ($\Gamma_1|0(\phi)\rangle_{\rm ABS}=0$) to the first excited state $|1(\phi)\rangle_{\rm {ABS}}=\Gamma_1^{\dagger}|0(\phi)\rangle_{\rm ABS}$ is given by the rate 
\begin{equation}
\label{pois:app:eq1}
W_{10}=\frac{2\pi}{\hbar}\sum_{if}\rho_i |\langle f|H_T|i\rangle |^{2}\,\delta(E_i-E_f).
\end{equation}
Here, the possible initial states are $|i\rangle=|0(\phi)\rangle_{\rm ABS}|\phi_i\rangle_{\rm lead}$ where $|\phi_i\rangle_{\rm lead}$ is the initial state of the dissipative channel coupled to a reservoir at temperature $T$ having the same chemical potential $\mu$ as the superconductors, and $\rho_i$ is the probability with which this initial state occurs. The possible final states are determined by $H_T$ and have the form $|f\rangle=c_{Rk\downarrow}^{\dagger}|1(\phi)\rangle_{\rm ABS}|\phi_i\rangle_{\rm lead}$ and $|f\rangle=c_{Rk\uparrow}|1(\phi)\rangle_{\rm ABS}|\phi_i\rangle_{\rm lead}$. Note that the fact that the process can create or annihilate an electron in the dissipative channel is a consequence of the ABSs being superpositions of electrons and holes. Both processes create a spin up excitation in the Fermi sea of the dissipative channel. The initial and final energies of the total $(L,R)$-system are $E_i=\varepsilon_{\phi_i} - \varepsilon(\phi)/2$ and $E_f=\varepsilon_{\phi_i}\pm \varepsilon_{k\downarrow/\uparrow}+\varepsilon(\phi)/2$. Using $\sum_{i}\rho_i\langle \phi_i|c_{Rk\sigma}^{\dagger}c_{Rk\sigma}|\phi_i\rangle\equiv {\rm Tr}[\rho c_{Rk\sigma}^{\dagger}c_{Rk\sigma}]=f(\varepsilon_{k\sigma})$ with $f(x)=[1+\exp(\beta x)]^{-1}$ the Fermi function in the dissipative channel, and $1-f(x)=f(-x)$, we obtain for the total rate
\begin{equation}
\label{pois:app:eq2}
W_{10}=\gamma_{\rm t}\varphi^{2}(0)f(\varepsilon(\phi)),
\end{equation}
with the tunneling rate per length $\gamma_{\rm t}=2\pi\nu|t|^2/\hbar$ where $\nu=\sum_k\delta(\varepsilon_{k\sigma}-\varepsilon)/{\bar l}$ is the (constant) density of states per spin and length at energy $\varepsilon$ counted from $\mu$. 

A similar calculation holds for the opposite rate $W_{01}$ where a spin down excitation tunnels into the dissipative channel with the result
\begin{equation}
\label{pois:app:eq3}
W_{01}=\gamma_{\rm t}\varphi^{2}(0)(1-f(\varepsilon(\phi)).
\end{equation}

    \end{appendix}

	%

\end{document}